# Improved Confinement in JET High β Plasmas with an ITER-Like Wall

C D Challis[1], J Garcia[2], M Beurskens[1], P Buratti[3], E Delabie[4], P Drewelow[5], L Frassinetti[6], C Giroud[1], N Hawkes[1], J Hobirk[5], E Joffrin[2], D Keeling[1], D B King[1], C F Maggi[1], J Mailloux[1], C Marchetto[7], D McDonald[1,8], I Nunes[9], G Pucella[3], S Saarelma[1], J Simpson[1] and JET contributors*

*EUROfusion Consortium, JET, Culham Science Centre, Abingdon, OX14 3DB, UK*
[1]*CCFE, Culham Science Centre, Abingdon, OX14 3DB, UK*
[2]*CEA, IRFM, F-13108 Saint-Paul-lez-Durance, France*
[3]*Unità Tecnica Fusione, C.R. ENEA Frascati, CP65, 00044 Frascati, Italy*
[4] *FOM-DIFFER, P.O. Box 1207 NL-3430 BE Nieuwegein, The Netherlands*
[5]*Max-Planck-Institut für Plasmaphysik, D-85748 Garching, Germany*
[6]*VR, Fusion Plasma Physics, KTH, SE-10044 Stockholm, Sweden*
[7]*Instituto di Fisica del Plasma, CNR, 20125 Milano, Italy*
[8]*EUROfusion PMU Garching, D-85748 Garching, Germany*
[9]*Instituto de Plasmas e Fusão Nuclear, IST, Universidade de Lisboa, Portugal*

Email address of main author: *clive.challis@ccfe.ac.uk*

**Abstract.** The replacement of the JET carbon wall (C-wall) by a Be/W ITER-like wall (ILW) has affected the plasma energy confinement. To investigate this, experiments have been performed with both the C-wall and ILW to vary the heating power over a wide range for plasmas with different shapes. It was found that the power degradation of thermal energy confinement was weak with the ILW; much weaker than the IPB98(y,2) scaling and resulting in an increase in normalised confinement from $H_{98}$~0.9 at $\beta_N$~1.5 to $H_{98}$~1.2-1.3 at $\beta_N$~2.5-3.0 as the power was increased (where $H_{98}=\tau_E/\tau_{IPB98(y,2)}$ and $\beta_N=\beta_T B_T/aI_P$ in %T/mMA). This reproduces the general trend in JET of higher normalised confinement in the so-called 'hybrid' domain, where normalised β is typically above 2.5, compared with 'baseline' ELMy H-mode plasmas with $\beta_N$~1.5-2.0. This weak power degradation of confinement, which was also seen with the C-wall experiments at low triangularity, is due to both increased edge pedestal pressure and core pressure peaking at high power. By contrast, the high triangularity C-wall plasmas exhibited elevated $H_{98}$ over a wide power range with strong, IPB98(y,2)-like, power degradation. This strong power degradation of confinement appears to be linked to an increase in the source of neutral particles from the wall as the power increased, an effect that was not reproduced with the ILW. The reason for the loss of improved confinement domain at low power with the ILW is yet to be clarified, but contributing factors may include changes in the rate of gas injection, wall recycling, plasma composition and radiation. The results presented in this paper show that the choice of wall materials can strongly affect plasma performance, even changing confinement scalings that are relied upon for extrapolation to future devices.

## 1. Introduction

The replacement of the JET carbon wall by an ITER-like wall, composed of beryllium for the main chamber wall and tungsten in the divertor, has affected plasma energy confinement in two ways: (i) by the effect of the wall materials on key plasma parameters, e.g. through plasma composition and wall recycling; and (ii) by operational techniques necessary to avoid damage to plasma facing components (PFCs) and/or maintain stable plasma conditions, e.g. deuterium gas injection to avoid tungsten accumulation. This has generally resulted in a reduction in confinement for ELMy H-mode plasmas in the ITER 'baseline' domain ($\beta_N$ <2) compared with previous C-wall experiments and the IPB98(y,2) confinement scaling [1,2]. By contrast, 'hybrid' plasmas at high $\beta_N$ exhibit similar confinement quality with the carbon and metal PFCs, significantly exceeding the IPB98(y,2) scaling in both cases [3]. These observations have motivated an experiment to investigate the scaling of energy confinement with applied heating power with the JET ILW and compare the results with previous experiments performed with the C-wall.

The observation of improved confinement with respect to the IPB98(y,2) scaling at high β has been reported on several tokamak devices, typically associated with an optimisation of the q-profile shape in plasma scenarios variously described as 'improved H-mode', 'advanced inductive' or 'hybrid' [e.g. 4-6]. The JET experiments used a 'current overshoot' technique before the main heating pulse to transiently shape the q-profile, suggesting that this was an important factor in the achievement of $H_{98}>1$ [6]. However, a subsequent comparison with experiments with a fully relaxed q-profile at the time of the main heating have shown that heating plasmas to higher β can result in elevated values of $H_{98}$ even when the q-profile shape is not optimised [7]. Furthermore, H-mode experiments to vary β while keeping other key dimensionless plasma parameters (such as normalised gyroradius, collisionality and safety factor) constant have also shown a much weaker β scaling of confinement than that suggested by the IPB98(y,2) scaling (i.e. $B\tau_E \sim \beta^0$ rather than $\beta^{-0.9}$) [8,9]. Taken together, these results suggest that power degradation of confinement can be much weaker than that indicated by the IPB98(y,2) scaling, even when the q-profile changes commonly employed in plasma scenarios at low and high β are avoided. To investigate the power scaling of confinement separately from the effects of the q-profile on plasma transport processes, the analysis reported in this paper was made for experiments where the q-profile shape was not varied.

It is well known that plasma confinement is sensitive to the plasma shape (e.g. [10]) and analysis of the JET C-wall experiments has shown that changes in plasma shape can also modify the scaling of energy confinement with power. It is believed that this effect was due to the source of neutral particles resulting from plasma-wall interactions in the main chamber [11]. Consequently the ILW experiments were performed using two different plasma shapes in order to investigate the impact of the change of wall materials on the shape dependence of confinement. This has resulted in a C-wall and ILW database comprising four power scans performed with two different plasma shapes and two different PFC arrangements. The confinement analysis of this combined set of experiments is the subject of this paper.

2. Overview of power scan experiments

The time evolution of typical plasmas from the JET C-wall and ILW power scan experiments are illustrated in Fig.1. The plasma current waveform shows a 'current overshoot' before the main auxiliary heating pulse provided by Neutral Beam Injection (NBI). This feature of the plasma current waveform forms a wide region of low magnetic shear in the plasma interior with $q_0$ close to unity, which has been developed at JET to provide access to the 'hybrid' domain with high $\beta_N$ and $H_{98}>1$ [6]. Thus the experiments described in this paper are representative of JET 'hybrid' plasmas at high power and, in terms of $\beta_N$, overlap with the 'baseline' domain at low power.

Power scans have been performed in plasmas of this type using two different plasma shapes, classified as having low and high triangularity (δ), respectively. Typical examples from the ILW power scans are illustrated in Fig.2. The plasma shape was influenced by plasma parameters that were varied during the power scans, such as poloidal β, which resulted in a weak collinear increase in triangularity as the heating power was increased. The result of this effect is illustrated in Fig.3, where the upper triangularity is plotted as a function of the heating power absorbed by the plasma for the four power scans. It can be seen that the shape match between the C-wall and ILW experiments was quite good in terms of upper triangularity and that the difference in this shape parameter between the low $\delta$ and high $\delta$ power scans was much larger that the variation in each of the individual scans.

Although the upper triangularity was quite well matched with the two sets of wall materials, the plasma configuration in the divertor was changed between the C-wall and ILW experiments described in this paper. Fig.4 shows the divertor configuration for typical plasmas from each of the four power scans. The ILW power scans had the outer divertor strike-point on the outer edge of the horizontal tungsten tile in the middle of the divertor and the inner strike-point on the tungsten coated vertical tile. The high triangularity C-wall power scan had a similar divertor configuration except that the outer strike-point was on the outermost horizontal tile, closer to the cryopump. For the C-wall low triangularity experiments a different divertor configuration was used with both inner and outer strike-points in the corners of the divertor. The reason for this change in divertor configuration was that the C-wall experiments were performed with configurations optimised for pumping whereas, at the time of the ILW experiments, routine operation had not yet been established using the outermost tungsten coated horizontal tile and it was considered desirable to characterise the confinement properties of plasmas in the conditions used for the bulk of the ILW dataset up to that point. Subsequent ILW experiments using the same divertor configuration as the C-wall power scans showed that the small change in the high $\delta$ scans did not result in a significant change in confinement. At low $\delta$ the energy confinement was increased by ~10% when the divertor strike-points were moved to the positions used for the C-wall power scan. However, this effect was consistent over a range of power levels such that the power scaling of confinement was not affected by the change in divertor configuration. A more thorough investigation of the effect of divertor configuration on plasma confinement with the JET ILW is presented in [12].

The deuterium gas injection rate during the main heating phase did not change significantly within each power scan, but a higher gas flow rate was used in the ILW power scans compared with the C-wall experiments. This was due to the need for some minimum level of gas injection in the ILW experiments to avoid rapid plasma contamination by high Z impurities, which can cause radiation cooling and a collapse of the plasma stored energy. Although impurity accumulation can be mitigated by gas injection, the use of excessive gas flow rates can also degrade the plasma energy confinement [13]. Consequently, the gas injection rate for the ILW power scans was set as low as possible to provide good ELMy H-mode confinement while ensuring that any rise in plasma radiation was on a long timescale compared with the energy confinement time so as to allow a study of the effect of the change in wall materials on the plasma confinement with as few additional changes as possible.

The process of gas optimisation is illustrated in Fig.5, where the time evolution of three ILW high triangularity plasmas with different gas flow waveforms is compared. In pulse #84540 the gas flow rate was kept constant during the main heating phase, which resulted in a gradual rise in the D$\alpha$ emission, suggesting a gradual rise in the neutral deuterium density in the tokamak chamber. This increase is correlated with a gradual increase in ELM frequency and decrease in plasma energy confinement, illustrated by the $\beta_N$ evolution in Fig.5. The other two pulses illustrated in Fig.5 show that, by ramping the gas injection rate down during the NBI heating pulse, it was possible to achieve roughly steady levels of D$\alpha$ emission, ELM frequency and plasma confinement. It is clear from this data that the plasma energy confinement was progressively degraded as the gas flow rate was increased. Therefore the gas injection waveform illustrated for pulse #84789 was chosen for the ILW power scans to provide good quality confinement in steady conditions for sufficient duration to allow an assessment of the power scaling of confinement. It should be noted that the level of Dα emission was slightly different for pulses #84540 and #84789 in the initial phase of the main heating pulse, despite the similar gas injection rates at the time. This difference is probably due to slight changes in first wall recycling conditions on the two different days over which the experiment was performed. It can be seen from Fig.5 that the effect of this change on the energy confinement time was relatively small. Nevertheless, in order to prevent such factors from introducing systematic uncertainties in the confinement analysis, the points in the ILW high δ power scan, which was carried out over the two separate days, were interleaved so that a mixture of high and low power pulses were performed on each day. The gas injection rates used in the four power scans are shown in Fig.6. The values are illustrated at the times chosen for the confinement analysis in this paper, as specified later in this section. Although the flow rates used in the ILW power scans were higher than for the C-wall experiments, they were still relatively low compared with the typical values used in JET ILW experiments, especially in the low β 'baseline' ELMy H-mode domain [3].

The use of minimal gas injection during the main heating phase of the ILW power scans resulted in a slow impurity accumulation in some cases with an accompanying gradual rise in the radiated power loss from the plasma. This effect was most clearly seen in plasmas with low heating power. As the heating power level was increased the radiated power fraction was reduced and the steady rise in plasma radiation was mitigated. This is illustrated in Fig.7, which shows two plasma pulses from each of the ILW power scans. In the case of the low δ plasmas the ELM frequency was lowest at low power (indicated by the frequency of the upward 'spikes' on the total radiation trace) and the radiation continued to rise during the pulse, both in terms of the total radiation and the radiation originating from within the Last Closed Flux Surface (LCFS). In the higher power case the ELM frequency was higher, as is typical for type 1 ELMs, and the plasma radiated a smaller fraction of the heating power; this fraction remained roughly constant during the main heating phase. Surprisingly, the ELM frequency remained roughly constant with power in the high δ ILW experiment. But despite this unusual behaviour the characteristic features of the radiation at low and high power were similar to the low δ power scan, as seen in Fig.7.

The use of a gradually decreasing gas injection rate during the heating pulse in these ILW experiments may also have affected the time evolution of the source and transport of impurities, and it is unlikely that even the roughly steady radiated power fraction achieved in the high power plasmas could be maintained indefinitely. Nevertheless, for the purpose of this confinement study it was considered more relevant to approach, as closely as possible, the low gas injection rates used in the equivalent C-wall power scans so that the effect of the wall materials themselves could be investigated. The effect of this was to produce a downward trend for the radiated power fraction in the ILW experiments as the heating power was increased, which was not seen in the C-wall experiments. The implications of the different radiation trends for the analysis of the energy confinement are discussed in the next section.

Three of the four power scans discussed in this paper were performed at the same plasma current and magnetic field (1.4MA/1.7T). In these cases the start time of the NBI main heating pulse was kept constant at t=4s, although sometimes it was preceded by a short period of low power preheating, as shown in Fig.1. The confinement analysis time window (t=5.0-5.4s) was chosen to be after the plasma stored energy had become roughly steady while being close enough to the start of the NBI heating pulse to avoid significant variations in the q-profile shape due to power dependent factors, such as changes in the resistivity and bootstrap current profiles. The only exception was the highest power plasma in the ILW high $\delta$ power scan, where the analysis time window was set to t=5.0-5.2s to exclude the effects of a short ELM-free phase, which provided a transient increase in plasma confinement. Unlike the other three power scans the low triangularity C-wall experiments were performed at 1.7MA/2.0T because this matched the best reference plasmas available at the time. This power scan also used a current overshoot before the NBI heating, but the timing of the high power heating pulse was delayed relative to the other experiments to compensate for the longer time required to reach $q_0$ close to unity. In this case the NBI start time was kept fixed at t=4.8s and the analysis time window was set to t=6.5-7.0s. All four power scans were performed at similar values of $q_{95}$ (~3.9).

For all plasmas in this experiment the duration of the confinement analysis window was chosen to be equal to or longer than the energy confinement time. The q-profile shape at the time of the analysis window was determined using the EFIT equilibrium code [14] constrained using motional Stark effect (MSE), polarimeter, plasma pressure and external magnetic field measurements and is illustrated for each of the power scans in Fig.8. It can be seen that the four power scans had quite similar q-profile shapes, indicating that the current profiles obtained in the C-wall experiments were largely reproduced in the equivalent ILW power scans. Although it is close to the measurement uncertainty (roughly ±10%), the q-profile measurements shown in Fig.8 suggest that the C-wall plasmas may have had slightly higher values of q close to the plasma centre compared with the ILW experiments, especially in the case of the low $\delta$ power scan where $q_0$ appears to have been greater than unity. This is consistent with the observation of electron temperature sawteeth in low power plasmas just before the time of the analysis window in all power scans except for the low triangularity C-wall plasmas. The observation of steady confinement conditions in low triangularity C-wall plasmas up to and including the sawtoothing phase suggests that such small differences in the q-profile shape are unlikely to have had a significant effect on the overall plasma energy confinement provided that performance limiting MHD instabilities, such as 3/2 tearing modes, were not present. This is illustrated in Fig.9 where the time evolution of a low power plasma from the C-wall low triangularity power scan is shown. The energy confinement reached steady conditions before the time of the analysis window (t=6.5-7.0s) and remains steady up to, and beyond, the time of the first sawtooth, which can be seen as an abrupt drop in the central electron temperature at t=7.82s. At the time of the second sawtooth, at t=8.87s, a continuous n=2 tearing mode was triggered (probably a 3/2 tearing mode), which finally degraded the energy confinement. Since real-time feedback control was used in this pulse to maintain a roughly constant value of $\beta_N$, the degradation of confinement is partly observed as an increase in the level of heating power required to prevent the loss of plasma stored energy. The loss of confinement when the n=2 mode was present can also be seen as a reduction in the value of $H_{98}$. On the basis of such observations it is concluded that the variations in the q-profile shape in the experiments discussed in this paper are unlikely to introduce significant changes in the global energy confinement at the time of the windows chosen for the analysis. Tearing modes, on the other hand, do have a significant impact on the energy confinement in these experiments, as seen in Fig.9. For this reason plasma pulses with significant levels of performance degrading MHD instabilities during the analysis window were excluded from the analysis dataset. This allows an investigation of the effects of the wall materials on energy confinement in plasmas with a similar internal magnetic geometry.

At the same time as the installation of the ITER-like wall in JET the NBI system was upgraded. The beam accelerators were changed from a combination of 80keV and 140keV systems to a higher power arrangement with 125keV accelerators and ion sources that were modified to increase the fraction of the power carried by the half and third energy components of the beam [15]. This has resulted in a systematic change in the energy spectrum of the neutral beam particles used to heat C-wall and ILW plasmas, which is further complicated by the choice of accelerator voltages for these specific power scans that were typically below the nominal values noted above due to the system capabilities at the time of the experiments and the need to avoid excessive beam shine-through power loading on PFCs. To assess the impact of these factors on the experiments discussed in this paper, the NBI power deposition profiles have been calculated using the TRANSP simulation code [16] and have been plotted in Fig.10 for four plasmas with NBI heating in the range 10-11MW, one from each of the power scans. It can be seen that the power deposition profiles were comparable for the four cases illustrated in terms of both core heating and the fractions of heating power to thermal ions and electrons. Also shown in Fig.10 are the radial profiles of the volume integrated beam heating power absorbed by each of the four plasmas. The variations in the total power absorbed by each of the plasmas were mainly due to small differences in the injected beam power. The similarity in the shape of the integrated power profiles indicates that the local heat fluxes were roughly the same over the whole plasma radius for steady plasmas in the four power scans when comparable NBI heating power was applied.

In conclusion, the data from four power scans have been assembled to allow an analysis of the plasma energy confinement behaviour. The dataset includes two different plasma shapes (high and low triangularity) and two different sets of plasma facing components (C-wall and ILW). The four power scans were comparable in terms of q-profile shapes and heating power deposition. The divertor configurations were different for the C-wall plasmas compared with the ILW power scans in that the divertor strike-points were closer to the cryopump, which can result in reduced plasma density and improved energy confinement, but it is not thought that these differences significantly affect the scaling of confinement with power. The deuterium gas injection rate was also higher in the ILW experiments, which was required to control impurity accumulation. But the gas flow rate was minimised to mitigate, as far as possible, the associated degradation of confinement. The resulting dataset allows a comparison of the confinement behaviour with the JET C-wall and ILW where the most significant differences are the wall materials themselves and the effects of variations in pumping, gas fuelling and recycling. A confinement analysis using this dataset is presented in the next section.

## 3. Confinement analysis

The aim of the experiments described in this paper was to perform a set of power scans with all other key 'engineering' variables kept as constant as possible in each scan. For this purpose the 'engineering' variables considered were those used in the IPB98(y,2) confinement scaling (i.e. plasma current, toroidal magnetic field, line average electron density, plasma main ion mass, major radius, aspect ratio and elongation). Since the magnetic configuration of the plasma was imposed directly by control of the tokamak coil currents, the resulting variables such as plasma current, magnetic field, major radius, aspect ratio and elongation were all maintained constant within ±1.3% for each individual power scan. However, the plasma triangularity, not included in the selection of IPB98(y,2) scaling variables, did increase systematically with power within each scan due to the effect of increasing β, as shown in Fig.3. The main ion species was deuterium with typically less than 3% hydrogen concentration in all four power scans. No attempt was made to control the plasma density during these experiments in order to avoid any systematic variation of the gas injection rate. Despite this the line average electron density remained roughly constant, within ±5%, in each power scan, as shown in Fig.11. Two pulses were omitted from the confinement analysis dataset because they had higher density than was typical for the relevant power scan, although one of these was included in the pedestal analysis in the next section. It can be seen from Fig.11 that the low triangularity plasmas achieved lower density compared with the high triangularity cases, which is a typical observation for JET C-wall and ILW experiments. The achievement of the lowest density of all in the C-wall low $\delta$ experiments is consistent with the choice of divertor configuration for this power scan, which was optimised for pumping, as illustrated in Fig.4. The achievement of essentially constant conditions in terms of the key 'engineering' variables listed above during each of the power scan experiments allows a direct measurement of the dependence of plasma energy confinement on heating power in the different wall environments.

The thermal plasma stored energy can be determined in two different ways in JET. The first method is to use a measurement of the total plasma stored energy (e.g. from diamagnetic loop measurements) and subtract the contribution of the fast ion stored energy due to neutral beam and radio frequency heating, as determined using a plasma simulation code such as TRANSP. This technique was used to provide JET data for the database constructed to develop the IPB98(y,2) confinement scaling and, therefore, is the preferred method for evaluating $H_{98}$. An alternative method is to map measured profiles of plasma electron and ion temperature and electron density onto magnetic flux surfaces using an equilibrium reconstruction and then integrate the plasma pressure to evaluate the total thermal stored energy. For the experiments described in this paper the total stored energy from the diamagnetic loop measurements was compared with that calculated using the kinetic method plus 1.5 times the perpendicular fast ion stored energy from TRANSP simulations. When averaged over the analysis time window these quantities agree within ±10% for all plasmas in the four power scans, as shown in Fig.12. Given this good agreement the kinetic profile technique was used for the following analysis to allow a more detailed study of the contribution of different parts of the plasma (e.g. the H-mode pedestal or pressure profile peaking) to the overall confinement behaviour.

In most cases the electron temperature and density profiles were measured using a high resolution Thomson scattering system. In the absence of an absolute calibration for this instrument, the density profile was scaled to match the line integrated measurements made using a far-infrared interferometer during the main heating phase of the plasma pulse. The exceptions are five pulses from the C-wall low $\delta$ power scan where high resolution Thomson scattering data were unavailable and were substituted by data from the LIDAR system. The ion temperature and toroidal rotation profiles were taken from charge-exchange measurements. Charge-exchange measurements near the edge of the plasma suggest that the ion and electron temperatures were similar at the top of the H-mode pedestal. Since the temperature pedestal shape was measured more precisely for the electrons than the ions the ion temperature was set equal to the electron temperature for the simulation in this region. $Z_{effective}$ was assumed to be constant with radius and was taken from visible bremsstrahlung measurements, with the plasma composition being characterised by a single impurity ion assumed to be carbon for the C-wall experiments and beryllium with the ILW. The q-profile used in the simulations was taken from EFIT equilibrium reconstructions as described in section 2 and the plasma radiation profile was taken from tomographic reconstructions of 2-D bolometer camera data. The absorbed heating power used in the confinement analysis consists of the Ohmic heating power, calculated using TRANSP, and the NBI power, estimated taking into account shine-through losses calculated by TRANSP and charge-exchange losses estimated using the simple formula described in [17].

Examples of the electron and ion temperature and electron density profiles are illustrated at two power levels for each of the four power scans in Fig.13. With the exception of the C-wall high δ plasmas the main effect of increasing the heating power was to increase the temperature across the whole plasma radius, especially for the ions. In the C-wall high δ plasmas the temperature rise was negligible in the plasma periphery, although some increase in temperature peaking is evident. This resulted in a weaker increase in plasma stored energy as the power was increased for the high δ C-wall plasmas compared with the other three cases, as illustrated in Fig.14 which shows the plasma thermal stored energy as a function of the absorbed heating power for each of the power scans. The stored energy expected from the IPB98(y,2) scaling is also shown in each case and $H_{98}$ can be estimated from the ratio of the measurements to the IPB98(y,2) curves. It is immediately noticeable that both of the ILW power scans and the C-wall scan at low δ all exhibit a substantially faster increase in plasma stored energy than is expected from the IPB98(y,2) scaling, leading to an increase in $H_{98}$ above unity at high power. This can be described as a much weaker power degradation of thermal energy confinement in the experiment than that given by the scaling. The exception is the C-wall high δ power scan, which has $H_{98}>1$ at all levels of power. However, unlike the other power scans, the thermal stored energy increases slowly as the power is increased following a similar power dependence to the IPB98(y,2) scaling. Since all of these power scans were performed at fixed plasma current, magnetic field and plasma shape, the increase in thermal stored energy results in a corresponding increase in β. In this sense the result shown in Fig.14 is consistent with the general observation in JET ILW experiments that $H_{98}$ tends to be higher in 'hybrid' plasmas at high $β_N$ compared with 'baseline' ELMy H-mode plasmas in the domain $β_N <2$. The observation of stronger power degradation of confinement in the C-wall high triangularity plasmas is dissimilar to both the ILW power scan using the same plasma shape and the low δ experiments using the same wall. This suggests that some special factor is responsible for the power dependence in this case.

The change in the wall material in JET has resulted in changes in the plasma impurity composition and, consequently, radiation losses. Fig.15 shows the radiated power fraction as a function of absorbed heating power for the four power scans. It shows that the radiated fraction of the heating power remained roughly constant with power in the C-wall high δ plasmas and increased gradually with power, although at a much lower level, in the C-wall low δ experiments. By contrast, the radiated power fraction decreased significantly as the power was increased for both of the ILW power scans, as described in the previous section. This should not be taken as a general observation for the JET ILW experiments as the radiation in the confined plasma depends on the accumulation of impurities, which in turn depends on various factors such as gas injection rate, core MHD behaviour, level of central heating, etc. Nevertheless it raises the question as to whether the systematic change in radiation behaviour in this particular experiment played a significant role in the observed differences in the scaling of plasma stored energy with heating power with the different wall materials.

In Fig.16 the thermal stored energy is plotted as a function of the net loss power for the four power scans in this experiment. Here the net loss power is defined as the difference between the absorbed heating power and the power radiated from within the LCFS. It can be seen that the substantial change in the dependence of plasma stored energy on power between the C-wall and ILW in the high triangularity power scans is not reconciled when the radiated power is taken into account. The scaling of stored energy for the other three power scans also remain comparable when evaluated using the net loss power. From these observations it is clear that neither the weak power degradation of confinement with the ILW nor the differences with respect to the high triangularity C-wall power scan can be explained by the changes in radiation behaviour from within the confined plasma due to the use of different wall materials.

Given the substantial discrepancy in the power degradation of confinement observed in the JET ILW experiments compared with the IPB98(y,2) scaling, it is interesting to investigate whether this was due to a strong increase in the pressure at the top of the H-mode pedestal near the plasma edge or a reduction in the core transport. Local core confinement improvements, called internal transport barriers, have been seen on many tokamaks and can be triggered or increased by increasing the level of auxiliary heating power [18,19]. On the other hand, there is experimental evidence that the pedestal pressure can increase with increasing β when the heating power is varied [20]. To investigate the contribution of the H-mode pedestal to the overall energy confinement, the pedestal pressure contribution to the total thermal stored energy in the experiments has been estimated for the plasmas in three of the four power scans. The C-wall low δ power scan was excluded because high resolution Thomson scattering data was not available for the majority of plasmas in this case. For the purpose of this analysis the pedestal top was taken to be at the radial point corresponding to 89% of the square-root of the normalised toroidal flux ($\rho_{tor}$), as illustrated in Fig.13. It can be seen that the pedestal width does not vary significantly in terms of this radial coordinate. The contribution of the pedestal pressure, $W_{ped}$, was then evaluated by integrating an energy density profile over the plasma volume where the profile was set equal to the measured energy density in the pedestal region and equal to the pedestal top energy density inside the pedestal top radius. In this calculation the ion temperature was again assumed to be equal to the electron temperature in the pedestal region. The contribution of the plasma core to the thermal stored energy has been characterised by the core energy, $W_{core}=W_{th}-W_{ped}$, where $W_{th}$ is the thermal stored energy.

The resulting values of pedestal energy and core energy are shown as a function of absorbed power in Fig.17. This shows that the pedestal energy increased only very weakly with power for the C-wall power scan at high triangularity, confirming that the pedestal behaviour was a key factor in the strong power degradation of energy confinement in this case compared with the other three power scans. The two ILW power scans, however, exhibited a stronger increase in pedestal energy with heating power. But it was significantly weaker than the scaling of the overall thermal stored energy shown in Fig.14 and, in the case of the low δ plasmas, begins to approach the dependence of the IPB98(y,2) confinement scaling. In all the experiments discussed in this paper the core energy increased more rapidly with power than the pedestal energy. This is illustrated for three of the scans in Fig.17 and indicates that the confinement in the plasma core contributed significantly to the weak power degradation observed with the ILW.

To show the contribution of the electron density profile to this rapid increase in core energy with heating power, the density peaking factor has been evaluated for the power scans shown in Fig.18. For this purpose the peaking factor is defined as the ratio of the values at $\rho_{tor}$=0.31 (representing the plasma core) and $\rho_{tor}$=0.89 (representing the plasma pedestal). The results are shown in Fig.18 where it can be seen that the electron density peaking behaved differently for the three power scans. It was essentially constant with power in the C-wall high triangularity plasmas, increased gradually with power for high δ ILW plasmas and rose most rapidly with power in the low δ ILW scan. This strong increase in density peaking for the low δ ILW plasmas tended to compensate for the weaker increase in pedestal energy with power compared with the high δ ILW experiments, as shown in Fig.17, to give the similar overall scaling of thermal confinement seen in Fig.14.

Fig.19 shows the electron and ion temperature peaking as a function of absorbed power for the power scans shown in Fig.14. Since temperature gradients can be steep close to the centre of the plasma, affecting the temperature profile peaking but not contributing significantly to the volume integral of the plasma pressure, the peaking factor is defined as the volume averaged temperature divided by the temperature at the pedestal so as to illustrate the contribution of the profile peaking to the average temperature. In the case of the ion temperature the volume average was taken from the simulations for the confinement analysis described above and edge charge-exchange measurements were used to determine the pedestal temperature. It can be seen that the electron temperature peaking remained roughly constant as the heating was increased in the ILW power scans, whereas the core ion temperature increased significantly with heating power, contributing to the rapid increase in plasma stored energy. In the case of the high δ C-wall experiments the measurements suggest that both the electron and ion temperature peaking increased gradually with heating power, providing a similar overall contribution to the increase in plasma stored energy.

Finally, it is worth noting that the calculation of the thermal stored energy is sensitive to the plasma ion composition, which was systematically different for the C-wall and ILW experiments [21]. Fig.20 show the line averaged $Z_{effective}$ values measured using visible bremsstrahlung for the plasmas in the four power scans. It can be seen that $Z_{effective}$ was systematically lower in the ILW experiments. For the purpose of this analysis it was assumed that the dominant impurity was carbon for the C-wall experiments and beryllium from the ILW experiments. Taking this into account the typical ion dilution, $n_i/n_e$, was 95-97% in the ILW experiments compared with 80-86% for the C-wall plasmas. This, together with the higher ion temperature peaking, resulted in a higher proportion of the plasma energy stored in the thermal ions in the high power ILW experiments compared with the equivalent C-wall high δ plasmas.

In conclusion, four power scans have been performed using two different shapes and two different set of first wall materials. The results of the confinement analysis can be summarised as follows:
   (1) Power degradation of confinement was weak in both of the ILW configurations and the C-wall low δ plasmas, much weaker than the IPB98(y,2) scaling, resulting in elevated $H_{98}$ only at high power and, therefore, high β;
   (2) This is consistent with the general observation of higher $H_{98}$ in high β 'hybrid' plasmas compared with low β 'baseline' plasmas in these configurations;
   (3) The rapid increase in plasma stored energy with heating power in these plasmas was due to a rise in both pedestal pressure and core pressure peaking;
   (4) By contrast, the high triangularity C-wall plasmas exhibited elevated $H_{98}$ over a wide power range with strong, IPB98(y,2)-like, power degradation;
   (5) The dissimilar behaviour of the C-wall high δ plasmas suggests that special factors specific to this combination of plasma shape and wall material affected the pedestal confinement in this case.
In the next section candidate explanations for these observations are discussed.

## 4. Modelling and interpretation of results

The weak power degradation of energy confinement with the ILW appears to have three key components linked to the behaviour of the electron density profile, the temperature profile and the H-mode pedestal. In the discussion below these are treated in turn.

*4.1 Density peaking*

In these experiments the electron density peaking was highest in the case of the low triangularity experiments and tended to increase with heating power. Since each power scan was performed at roughly constant line average density, the average plasma temperature increased with heating power and, consequently, the collisionality decreased as the power was increased. The lower density in the low triangularity plasmas compared with the high $\delta$ cases also tends to result in lower collisionality in these cases. Thus the observed variations in density peaking in this dataset are correlated with the plasma collisionality, consistent with previous observations in ASDEX Upgrade and JET [22]. This correlation is shown for the four power scans in Fig.21. The general consistency between the C-wall and ILW plasmas in Fig.21 is partly due to the inclusion of $Z_{effective}$ in this definition of collisionality, which is systematically higher for the C-wall plasmas as shown in Fig.20. It should be noted that the central fuelling due to neutral beam injection also increased with heating power in each of the scans and it was not possible to decouple the effects of collisionality and fuelling in these experiments. Nevertheless the contribution of electron density peaking to the rapid increase of plasma stored energy with heating power in three of the four power scans discussed in this paper appears to be consistent with the previously observed behaviour in a wider database on ASDEX Upgrade and JET.

*4.2 Temperature peaking*

Previous analysis of a large JET C-wall dataset showed a general trend for electron temperature peaking to decrease and ion temperature peaking to increase as collisionality was reduced in the low collisionality domain [23]. This is qualitatively consistent with the C-wall low $\delta$ profiles illustrated in Fig.13, which show an increase in $T_i/T_e$ in the plasma core as the power is increased and, consequently, the collisionality is reduced. The behaviour of the C-wall high $\delta$ power scan is once again exceptional, showing a gradual increase in both the ion and electron temperature in the plasma core. This atypical observation may be related to the pedestal behaviour, where it is noticeable that the temperature remained roughly steady as the heating power was increased. This unusual pedestal behaviour is discussed further below.

In the ILW power scans the electron temperature peaking remained roughly constant as the heating power was increased, whereas the core ion temperature increased with respect to the electrons. Since each power scan was performed at roughly constant density, the shape of the NBI power deposition profile did not vary significantly as the power was increased. The increase in plasma temperature with heating power, however, resulted in a gradual increase in the fraction of the NBI power that was absorbed by the plasma ions as the total power was increased. This is illustrated in Fig.22, which shows the fraction of the absorbed heating power provided to the plasma ions calculated using the TRANSP simulation code. It is possible, therefore, that the relative increase in ion heating may have contributed to the increase in ion temperature profile peaking with heating power seen in these experiments.

Various mechanisms have been put forward in the literature as potential explanations for increased ion temperature peaking in plasmas of this type. Plasma rotation and rotational shear are thought to play a role in the suppression of turbulence [18,19] and observations of reduced confinement have been reported when torque was reduced in 'hybrid' plasmas in DIII-D [24]. In the power scans described here the torque applied to the plasma by the neutral beams increased collinearly with the heating power, making it impossible to rule out plasma rotation effects in this case. However, an alternative explanation for the increase in ion temperature peaking may come from the increased fast ion pressure content, which is characteristic of JET 'hybrid' plasmas [25] and enhances the local pressure gradient. This has a stabilising effect on ITG turbulence as shown in electromagnetic gyrokinetic simulations both linearly [26] and nonlinearly [27]. Fig.23 shows the suprathermal stored energy for the plasmas in the four power scans calculated using the TRANSP simulation code. Again the strong correlation between heating power and the fast ion pressure could be consistent with a transport reduction of this sort.

The correlation between ion heating, applied torque and fast ion pressure in these experiments does not allow a simple discrimination between the various mechanisms discussed above. Therefore predictive simulations of the core electron and ion temperature have been performed using the TGLF quasi-linear transport code [28] to further investigate the factors that may play a role in the core temperature increase with power observed in these experiments. Previous TGLF simulations of JET C-wall 'hybrid' experiments achieved good agreement with measured temperature profiles when fast ion and electromagnetic effects were taken into account, but indicated a more limited impact of E×B flow shear on turbulence [29]. To investigate the behaviour of the core transport in the ILW power scans the low $\delta$ plasmas with different heating power levels illustrated in Fig.13 were selected for analysis. The modelling was performed in the same way to the previous C-wall simulations with the electron density profile taken from the experiment and the edge temperature being set equal to the measured value at $\rho_{tor}$=0.9. The comparison of the predicted ion and electron temperature profiles with the measured profiles are shown for the full simulations (including E×B flow shear, fast ion and electromagnetic effects) in Fig.24. It can be seen that the simulations capture the increase in both core temperature and core $T_i/T_e$ when the power is increased.

To investigate the relative importance of various effects in the modelling the TGLF simulations were repeated, firstly with the fast ion and electromagnetic effects removed, and then with the additional removal of the E×B effects. The results are shown for the same two plasmas in Fig.25. The contribution of E×B flow shear is seen to be small in both cases, consistent with the observation in [29]. The contribution of fast ion and electromagnetic effects are more important for the high power case where the fast ion pressure is much larger, as shown in Fig.23. This suggests that the rapid increase in fast ion pressure gradient as the power is increased may play a role in the core temperature increase observed in the ILW power scans. TGLF, however, is a quasilinear model that does not fully capture nonlinear effects. Nonlinear gyrokinetic simulations show that nonlinear turbulence effects may indeed play a role in these plasmas [30] and therefore the impact of fast ions and electromagnetic effects can be even higher.

Finally, simulations were performed to investigate the effect of the increased fraction of the neutral beam power absorbed by the thermal ions as the heating power was increased. To do this the full TGLF simulation for the high power plasma shown in Fig.25 was repeated with the same level of heating power, but with the ratio of ion and electron heating power of the low power plasma illustrated in Fig.25. The resulting profiles shown in Fig.26 indicate a reduced ion temperature in the plasma core when the fraction of heating to the ions is reduced to the level of the low power plasma. It has been noted above that the collisionality decreased as power was increased in the power scans presented in this paper, which is expected to result in a progressive decoupling of ions and electrons. Thus it can be seen that, in addition to the plasma transport effects, variations in heating and coupling of the ions and electrons can also be a factor in the observed increase in core $T_i/T_e$ at high power.

*4.3 H-mode pedestal*

It has been shown above that a key factor for the H-mode pedestal confinement is the role of neutral particles in the tokamak chamber due to the injected gas, pumping and recycling. In the ILW experiments described in this paper the deuterium gas injection rate was minimised to match as closely as possible the approach adopted in the previous C-wall power scans. Fig.27 shows the D$\alpha$ emission measured in the tokamak mid-plane as a function of absorbed power for the C-wall and ILW power scan experiments. This shows that the D$\alpha$ emission was roughly constant with power for all of the power scans except the C-wall high $\delta$ case, where a much higher level of emission is observed, which increased rapidly with power. It should be noted that the level of the D$\alpha$ emission is probably over-estimated for the ILW experiments due to reflections within the tokamak vessel. But even taking into this into account it is apparent that the C-wall high $\delta$ experiment was exceptional in terms of both the level of main chamber neutral deuterium particles and the collinear increase with power. This trend is consistent with a combination of the close proximity of the high $\delta$ plasma to the plasma facing components and the recycling characteristics of the carbon wall, and may play a role in the weak increase in plasma stored energy with power in this case, due to a progressive degradation of the pedestal confinement [11]. A more thorough analysis of this issue using additional diagnostics and an assessment of candidate physics explanations for the effect on confinement will be discussed in a separate publication. For the purpose of this analysis of confinement dependence on heating power it is considered that the C-wall high $\delta$ power scan is affected by the collinear variation in main chamber neutral particle population; therefore this scan is considered atypical in terms of energy confinement scaling compared with scans performed in more controlled conditions.

In [2] it was noted that the effects of gas injection on the H-mode pedestal structure with the ILW, including the beneficial effects of impurity seeding [31], are not fully understood. Indeed, the presence of significant levels of carbon in the C-wall plasmas, indicated by the $Z_{effective}$ measurements shown in Fig.20, mean that a similar mechanism could have contributed to the good confinement seen at low power in the high $\delta$ C-wall plasmas. However, it is interesting to investigate whether the pedestal energy increase with power in the ILW power scans presented here, where the gas injection rate has been minimised, is consistent with modelling of peeling-ballooning mode instabilities. The H-mode pedestal in the phase just before ELM events has been characterised in terms of width and height using hyperbolic-tangent fits to the high resolution Thomson scattering profiles. The ELITE code [32] was then used to find the point of marginal stability by assuming the measured density profile shape and scaling the pedestal temperature at constant pedestal width (in normalised polioidal flux coordinates) until the stability boundary was found.

The results are shown in Fig.28 where the measured pedestal pressure is compared with the modelled value at marginal stability for selected plasmas from the ILW power scans with different plasma shapes and heating power levels. It can be seen that the observed increase in pedestal pressure with heating power is reproduced by the modelling, as is the divergence of the pedestal pressure for the high and low δ plasma shapes as the power is increased. This indicates that the more rapid increase in pedestal energy with power in the high δ ILW plasmas compared with the low δ ILW power scan shown in Fig.17 is consistent with the peeling-ballooning analysis. It should be noted that the global plasma β measurements have been matched in the modelling analysis so that the beneficial effect of the increase in core plasma energy on pedestal stability is taken into account. The stability modelling shows that the rapid increase in pedestal pressure with power comes in two parts: an increase in α, the normalised pressure gradient; and an increase in the pedestal width (measured in normalised poloidal flux coordinates), as shown in [33]. For the plasmas illustrated in Fig.28 the modelled marginally stable α increases by ~50% from the lowest to highest β at high δ, whereas the increase is ~25% for the low δ plasmas. The remainder of the increase in the modelled pedestal height shown in Fig.28 is due to changes in the pedestal width. The electron temperature and density pedestal profiles are shown in Fig.29 for the ILW high and low δ plasmas at three different power levels, clearly illustrating the increase in pedestal width as the power was increased. This is qualitatively consistent with the observed correlation between the pedestal width and the local poloidal β, which has been found on several tokamaks, and the argument described in [34] that this may be linked with the onset of strong electromagnetic kinetic ballooning mode turbulence in the pedestal region, although local gyrokinetic analysis suggests that kinetic ballooning modes are not always unstable in the fully developed JET pedestal [35].

*4.4 Interdependencies*

It has been noted in [36] that the combination of a favourable dependence of the pedestal height on the Shafranov shift, which increases with the global poloidal β, and core profile 'stiffness' would lead to a virtuous cycle. In this paradigm increasing the heating power raises the core pressure, leading to a larger Shafranov shift and allowing a higher pedestal pressure. The higher pedestal pressure then allows yet higher core pressure, and hence higher poloidal β, continuing the cycle. The observations of an increase in the pedestal pressure gradient as β was increased in the ILW power scans, which manifests itself as an increase in pedestal temperature and an invariance, or even increase, in the core temperature peaking, suggests that such a positive feedback loop may play a role in the weak power degradation of confinement in these experiments. Combined core-edge modelling of similar plasmas supports this hypothesis [37].

The detailed observations and modelling discussed in this section suggest that, if such a virtuous cycle were at play within these plasmas, there are additional factors that can act to enhance it. It has been suggested that the correlation between the density peaking and collisionality, seen in H-mode plasmas on many devices and reproduced in these experiments, can be due an inward pinch associated with ITG turbulence [38]. In this case an increase in core temperature would act to increase the core density peaking, further increasing β. Similarly the increase in core electron temperature is predicted to result in an increase in the fast ion slowing down time, which increases the total plasma β further. The argument in [27] that this can also reduce core ITG transport to allow higher core temperature would add yet another virtuous cycle to the feedback system. Finally, the increase in core temperature is also predicted to decouple thermal ions and electrons and increase the fraction of the neutral beam fast ion heating absorbed by the thermal ions, which resulted in a further increase in the core temperature in the transport simulations above. Taken together, these processes provide a set of positive feedback loops, as illustrated in Fig.30, that could contribute to the rapid increase in plasma stored energy with applied heating power seen in these ILW power scan experiments.

*4.5 Consistency with the wider JET ILW dataset*

It has been noted above that the power scan experiments described in this paper have been performed with low levels of deuterium gas injection. Whereas this is a common characteristic of JET plasmas in the high β 'hybrid' domain, low β 'baseline' experiments are usually performed with much higher gas injection rates. Since it has been seen that the pedestal pressure can be influenced by the population of neutral particles in the tokamak chamber, the ILW low δ power scan was repeated using higher deuterium gas injection rates, more typical of 'baseline' plasma operation with the ILW. A full confinement analysis using kinetic profiles is not yet available for these more recent experiments, so the thermal plasma stored energy has been evaluated from the diamagnetic measurements, as described in section 3, but calculating the fast ion stored energy using the PENCIL Fokker-Planck simulation code [39]. The analysis time window was delayed for the experiments with the highest gas flow rates to take into account the longer period before the plasma confinement became stationary. The results are shown in Fig.31 for three ILW low δ power scans: the experiments described in this paper; and two further power scans using the same divertor configuration as the C-wall low δ experiments, as illustrated in Fig.4, but with two different, higher gas injection flow rates.

The power scans with different gas flow rates exhibit completely different scalings for the energy confinement dependence on power. Unlike the weak power degradation of confinement seen in the ILW experiments discussed in this paper, the power scan with the highest gas flow rate shows strong, IPB98(y,2)-like, power degradation. To investigate this further the pedestal stability analysis described above was repeated for selected plasmas from the experiment with high gas injection [40]. In this analysis the pedestal pressure rose rapidly with heating power in the peeling-ballooning modelling whereas the measured pedestal pressure showed only a weak increase across the whole power scan with the measured value of α falling well below the level of marginal stability from the simulation. The reason for this discrepancy between measurements and modelling is not yet understood, but the experiment serves to illustrate the important role played by the pedestal in affecting the overall confinement scaling.

To put the ILW power scan experiments in context with the wider JET ILW dataset, the 'low gas' and 'high gas' power scans illustrated in Fig.31 have been compared with a large dataset of JET ILW 'baseline' and 'hybrid' experiments from [3]. Fig.32 shows the data plotted as the thermal energy confinement time normalised to $\tau_{98} \times P^{0.69}$ as a function of heating power, where $\tau_{98}$ is the confinement time given by the IPB98(y,2) scaling. Dividing by $P^{0.69}$ removes the power dependence of the IPB98(y,2) scaling so that, if all other dependencies in the scaling accurately describe the plasmas in the dataset, the overall power scaling of the dataset can be deduced. The curve corresponding to $H_{98}=1$ has also been added to illustrate the dependence expected by the IPB98(y,2) scaling. In Fig.32 the large dataset has been divided into two groups: plasmas with $\beta_{N,dia}<2$ (typical of 'baseline' experiments with high gas flow rates); and $\beta_{N,dia}>2$ (typical of 'hybrid' experiments with low gas flow rates). It is interesting to note that, in this dataset, the 'baseline' plasmas typically have $H_{98}<1$ while the 'hybrid' plasmas typically have $H_{98}>1$. Data from the ILW low δ power scans with the low and high gas flow rates illustrated in Fig.31 are also plotted in Fig.32. The 'high gas' power scan overlays the low β 'baseline' data, consistent with the picture that power degradation of confinement is strong in this domain. The 'low gas' power scan exhibits low $H_{98}$ at low power, connecting to the 'baseline' domain. But the weak power degradation of confinement allows these plasmas to cross to the high β 'hybrid' domain at high power. This indicates that the weak power degradation of confinement observed for the ILW power scans in this paper is a key factor for the achievement of high $H_{98}$ in JET high β 'hybrid' plasmas. This is consistent with the observation in other JET experiments that similar confinement is obtained in 'baseline' and 'hybrid' plasmas, despite the different initial q-profile shapes, when other 'engineering' parameters, such as plasma current, magnetic field, heating power, gas injection rate, etc. are matched [7].

## 5. Conclusions

The four power scans studied in this paper show that tokamak first wall materials can affect plasma performance, even changing confinement scalings that are relied upon for extrapolation to future devices. It has been concluded that the C-wall high triangularity experiments, which were the only case to exhibit strong power degradation of confinement, are atypical of the confinement behaviour of plasmas with low deuterium gas injection rates in the JET tokamak. This exceptional case appears to have been affected by a source of neutral particles in the main chamber, which, together with the observation of strong power degradation of confinement in conditions of high gas injection, raise important questions concerning the role of neutral particles in plasma confinement.

The main conclusion of this work is that, in the presence of the ITER-like wall materials and the absence of high gas injection rates, the confinement degradation with heating power is weak, much weaker than the IPB98(y,2) scaling. The observed rapid increase in plasma stored energy with power is due to the increase in pedestal pressure and core pressure peaking. The pedestal pressure increase is consistent with peeling-ballooning modelling, suggesting the importance of β effects in pedestal stability. The density peaking increase is consistent with the previously observed correlation with collisionality. Modelling of the core heat transport suggests that the increase in temperature peaking may be linked to an increased fraction of the heating power to the ions, as well as electrostatic and fast ion effects.

It is worth noting that the various components responsible for the rapid rise in plasma stored energy are not necessarily independent. The increase in the core pressure results in an increase in the global β, which, in turn, can affect the pedestal stability. Conversely, the increase in pedestal pressure can affect the core pressure if core transport processes are sensitive to gradient scale lengths. Thus the confinement degradation with heating power may be determined by the complex interplay between a variety of physics mechanisms rather than the simple addition of core and edge factors. The analysis in this paper shows that the weak power degradation of confinement seen in the JET ILW experiments cannot be simply extrapolated to other plasma conditions or devices. Instead it is necessary to assess the underlying physics processes of importance and determine the nature of any virtuous cycles that may exist. For example, in the case of ITER the factors concerning strong ion heating and the increase in $T_i/T_e$ are not expected to be as important as in the experiments discussed in this paper. However, the dependence of α-particle production on the core plasma pressure and the resulting generation of fast ion pressure and plasma heating will need to be taken into account when considering the core-edge interplay in burning plasmas. Nevertheless, the potential impact of weak power degradation of confinement in tokamak plasmas is encouraging for the development of high gain fusion devices where heating power and plasma β are high.

**Acknowledgements**

*This work has been carried out within the framework of the EUROfusion Consortium and has received funding from the European Union's Horizon 2020 research and innovation programme under grant agreement number 633053 and from the RCUK Energy Programme [grant number EP/I501045]. To obtain further information on the data and models underlying this paper please contact PublicationsManager@ccfe.ac.uk. The views and opinions expressed herein do not necessarily reflect those of the European Commission.*

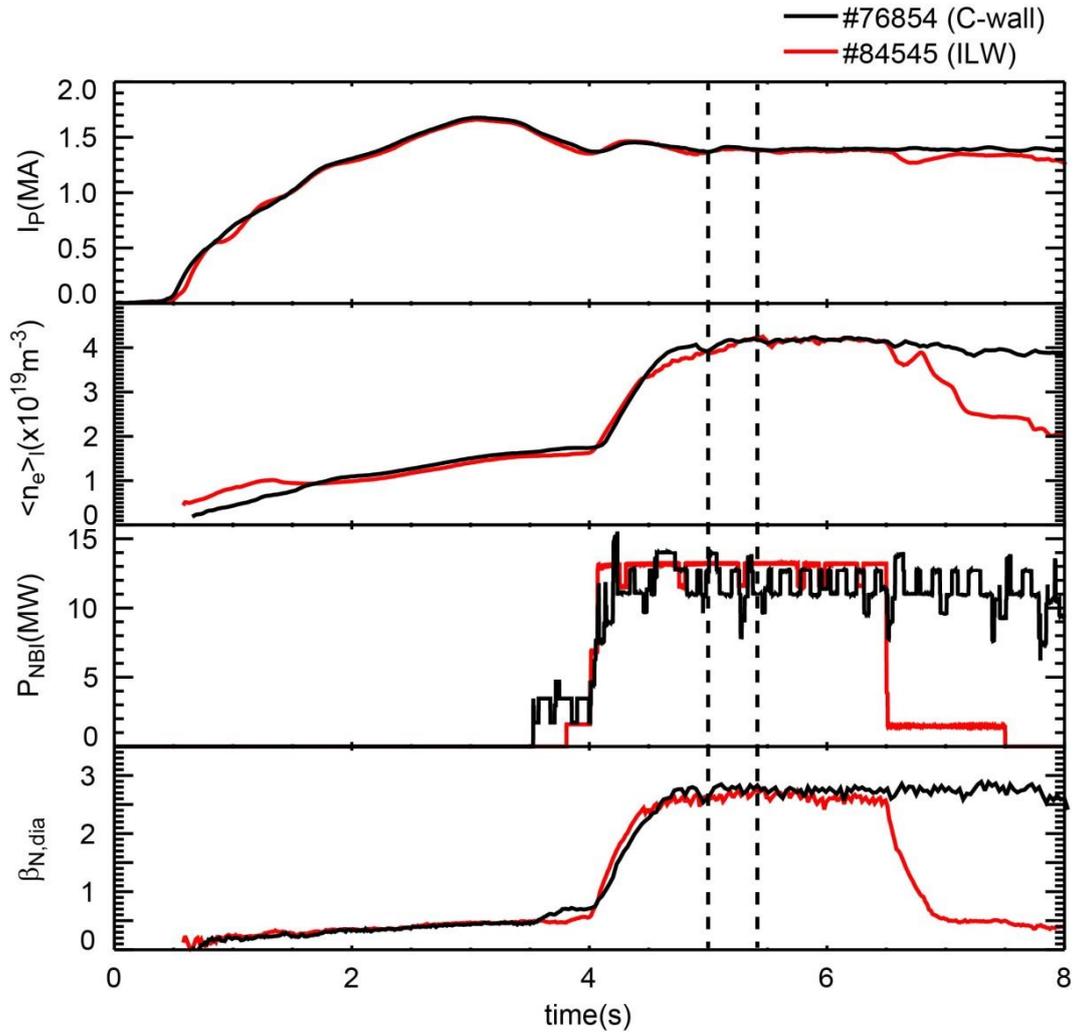

Fig.1. Time evolution of typical plasmas from the C-wall (black) and ILW (red) power scan experiments showing the plasma current, line averaged density from interferometry, auxiliary heating power and $\beta_N$ evaluated from the diamagnetic stored energy measurements. The two vertical lines indicate the time window used for the confinement analysis.

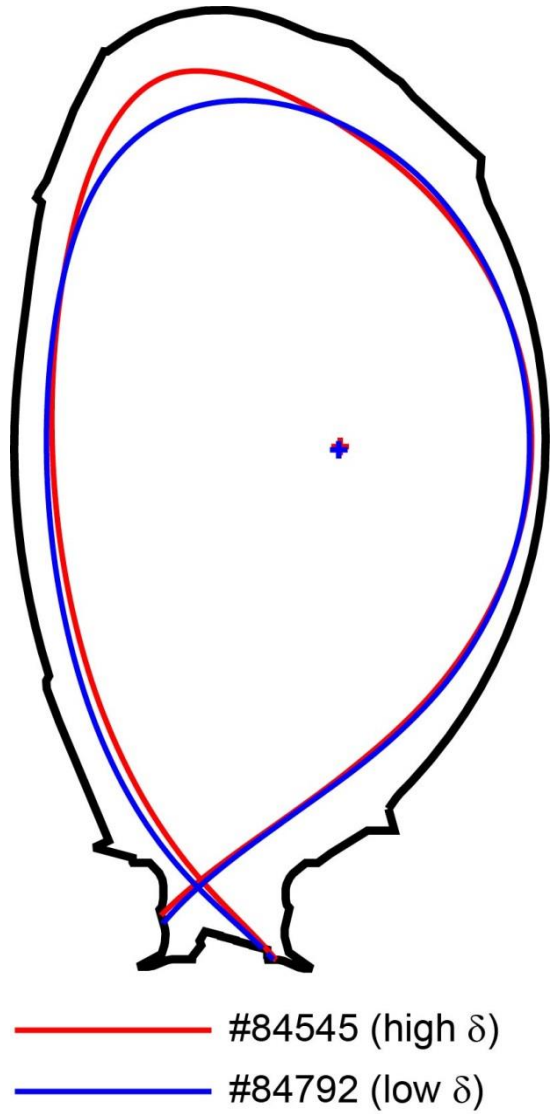

*Fig.2. Typical examples of high δ (red) and low δ (blue) plasma shapes used in the power scan experiments.*

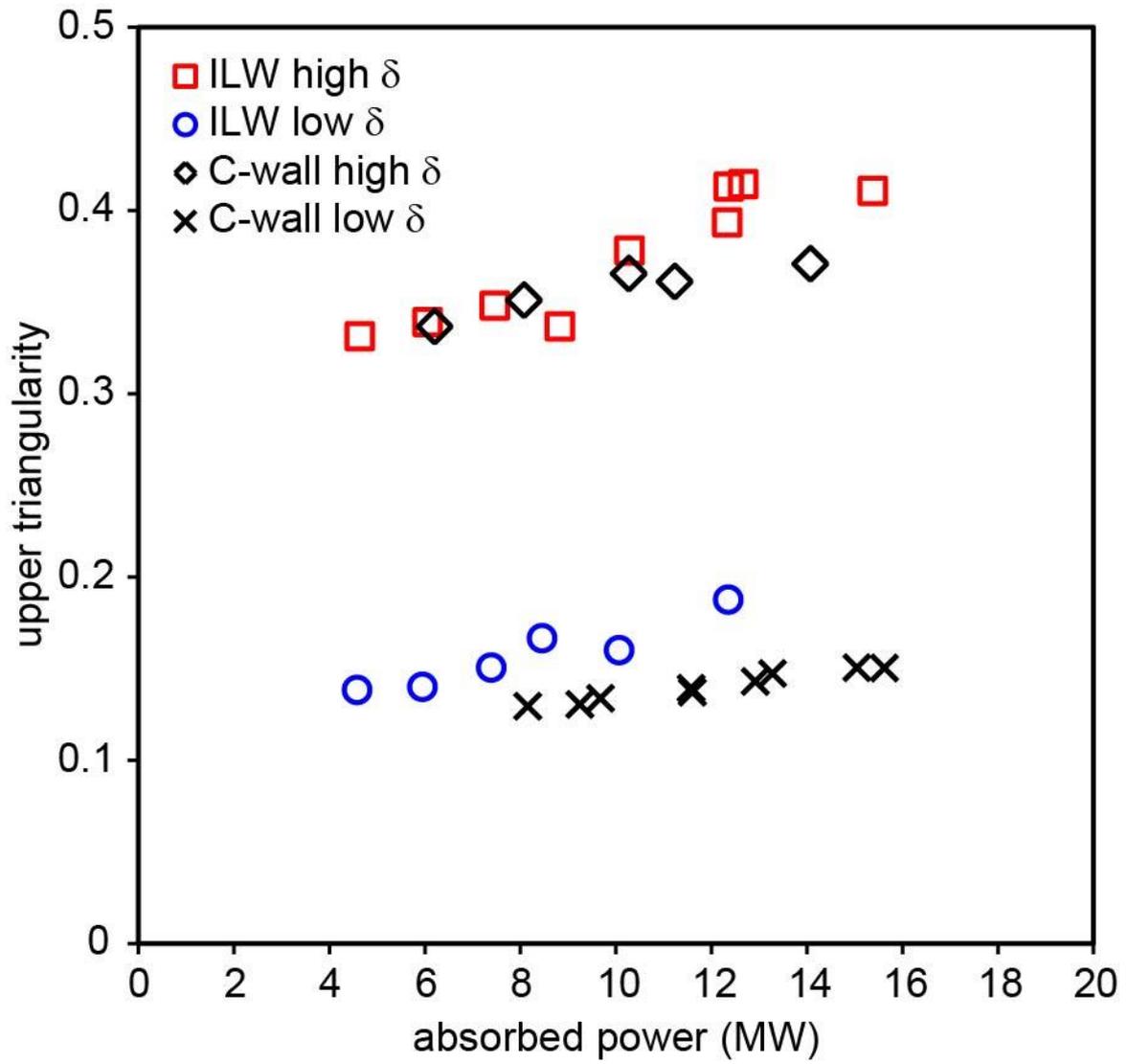

*Fig.3. Upper triangularity as a function of absorbed power for the four power scan experiments.*

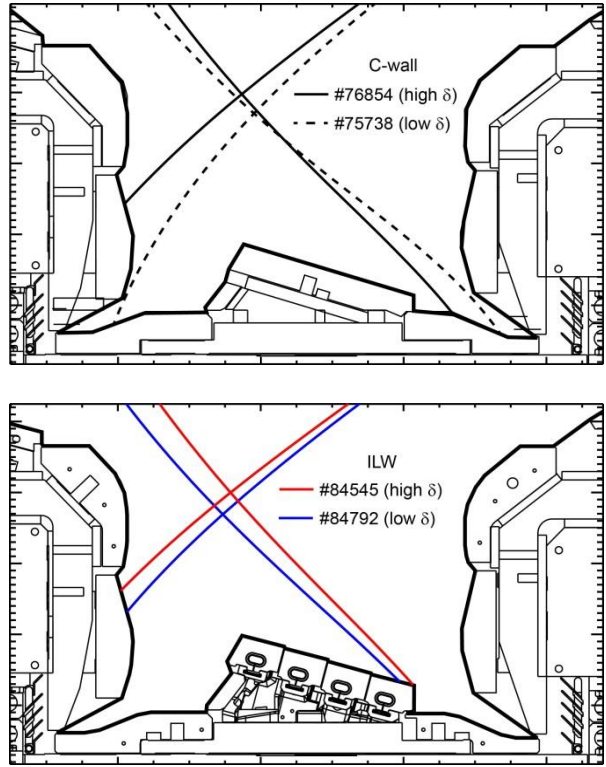

*Fig.4. Divertor configurations used for each of the four power scan experiments.*

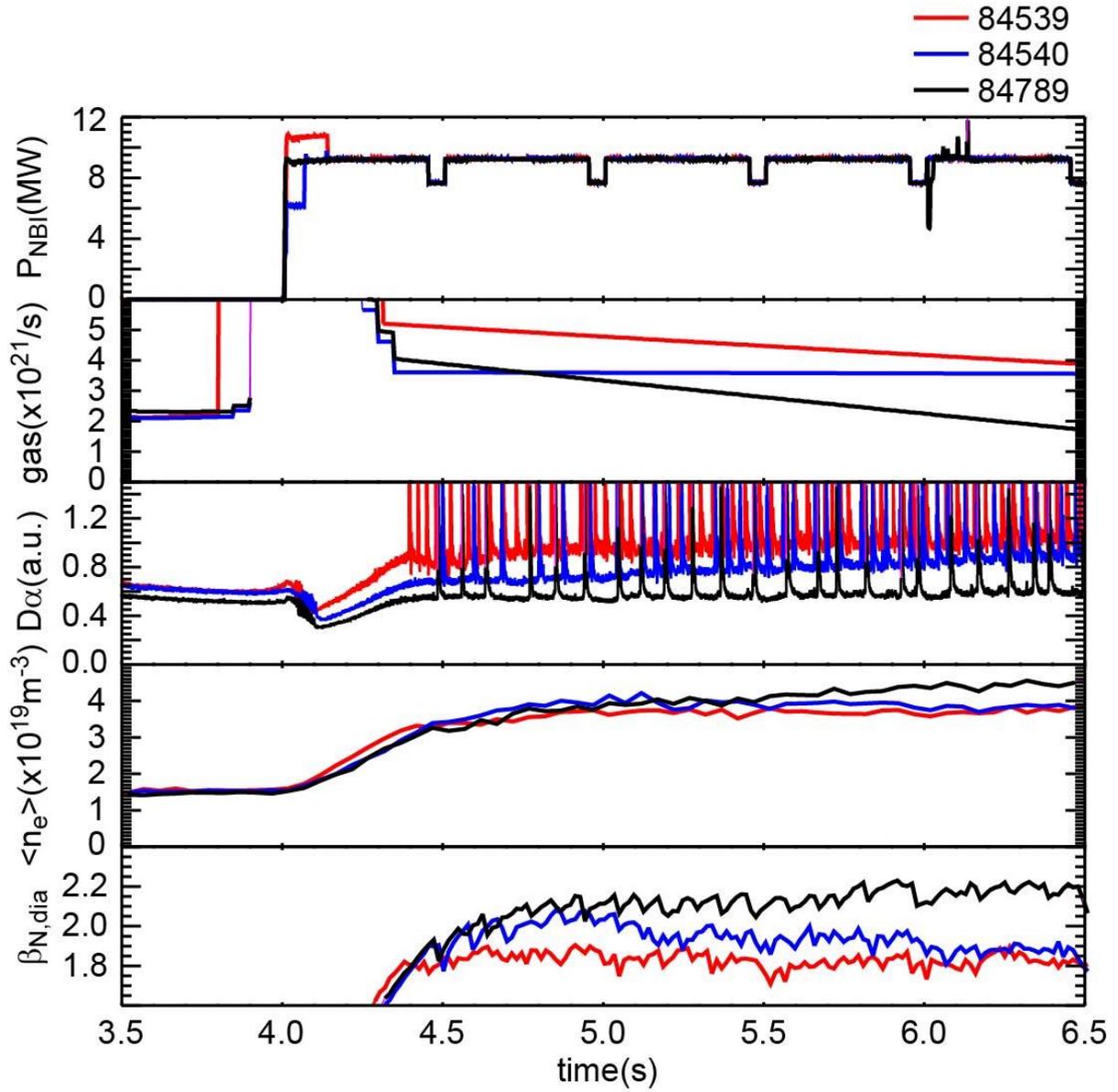

*Fig.5. Time evolution of NBI power, deuterium gas injection rate, line average density from high resolution Thomson scattering normalised to interferometer measurements, Dα emission from a line-of-sight that includes the outer region of the divertor and $\beta_N$ from the diamagnetic measurement for three ILW high $\delta$ plasmas at 1.4MA/1.7T with different gas injection waveforms.*

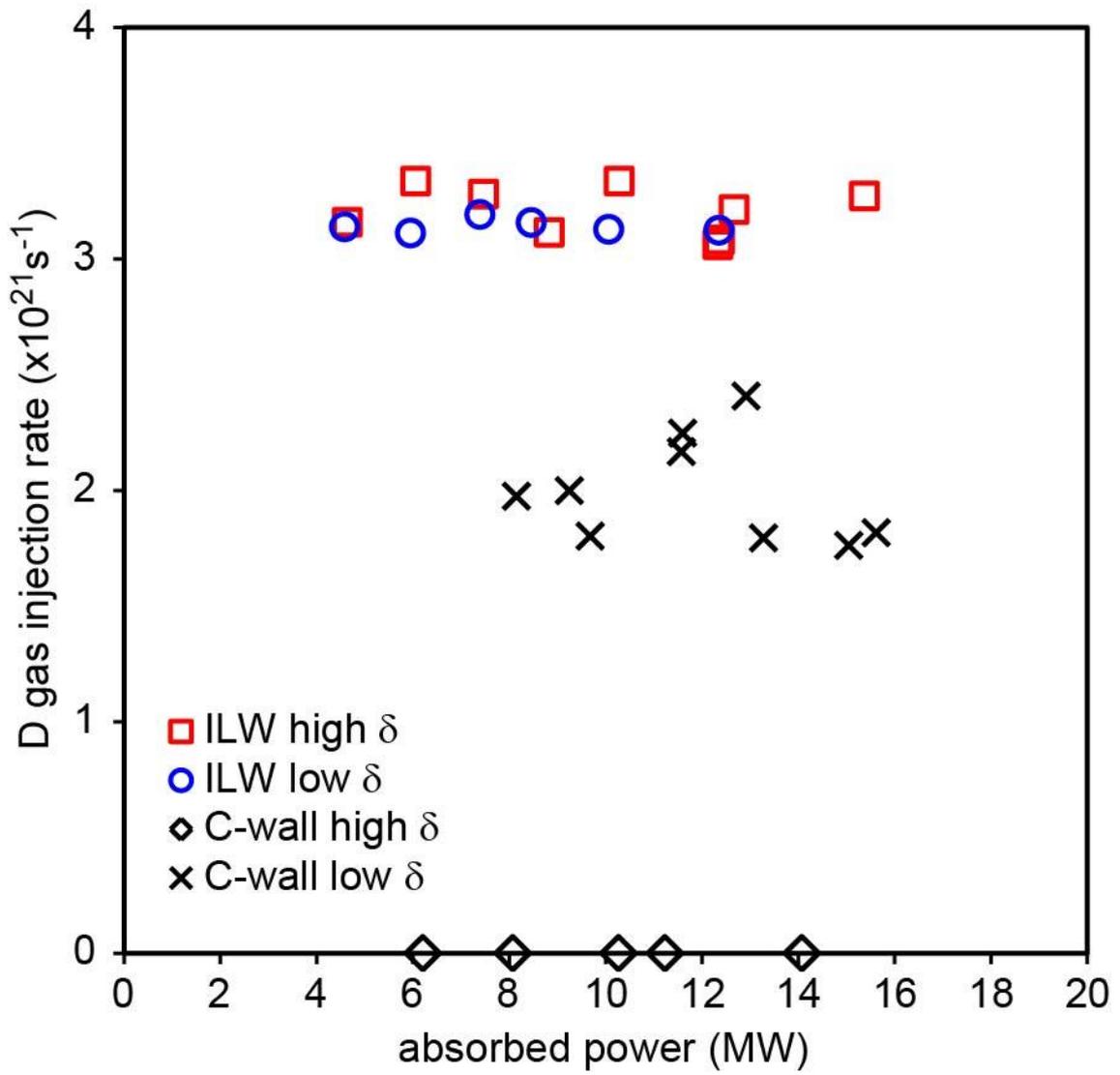

Fig.6. Deuterium gas injection rate during the main heating phase as a function of absorbed power for the four power scan experiments.

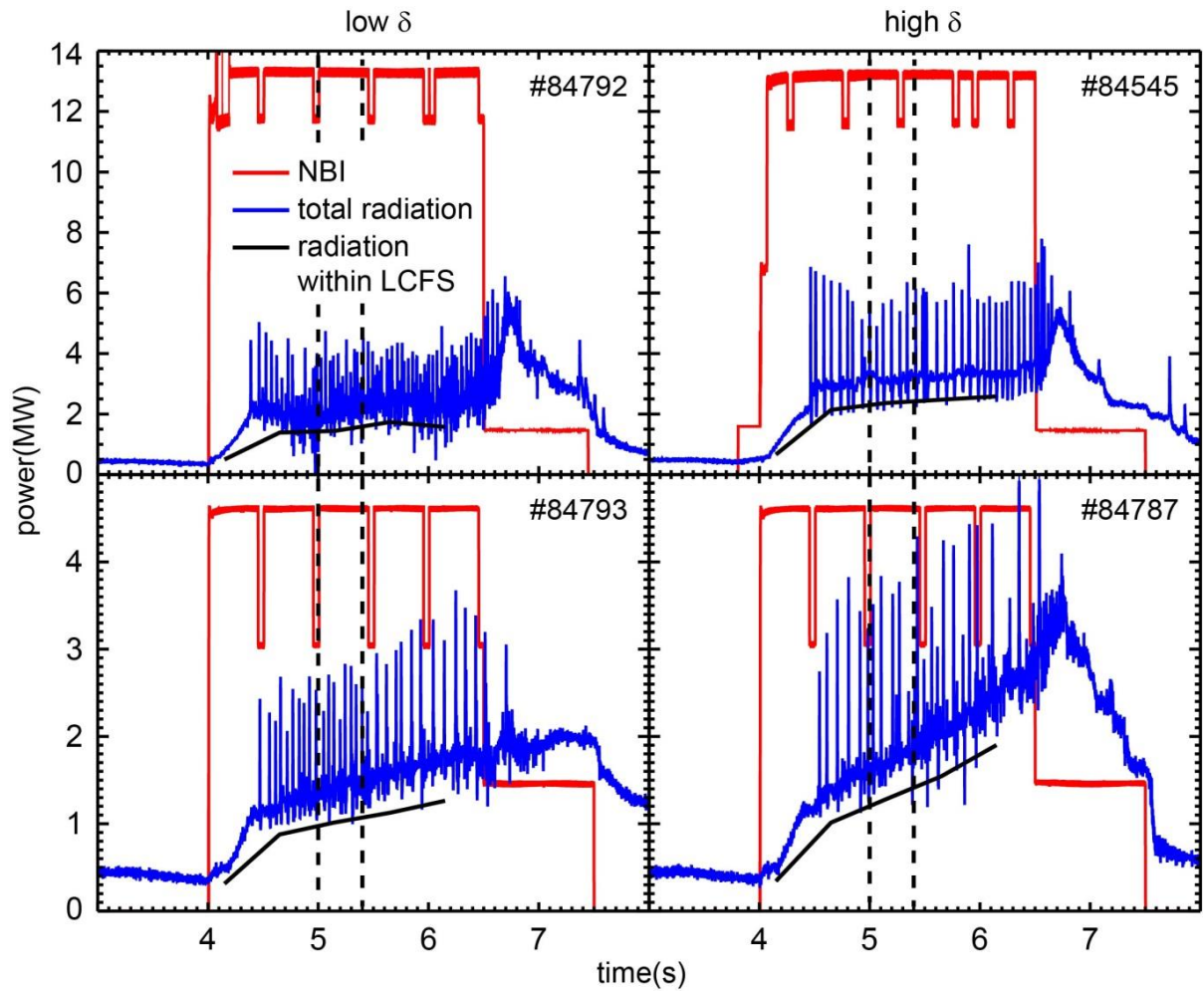

*Fig.7. Time evolution of NBI power, total radiation and radiation originating from within the LCFS for plasmas at two different power levels from the low δ (left) and high δ (right) ILW power scans. The total radiation was determined from bolometer measurements and the radiation from within the LCFS was evaluated using a tomographic reconstruction of the data from the 2-D bolometer camera. The pairs of vertical lines indicate the time window used for the confinement analysis.*

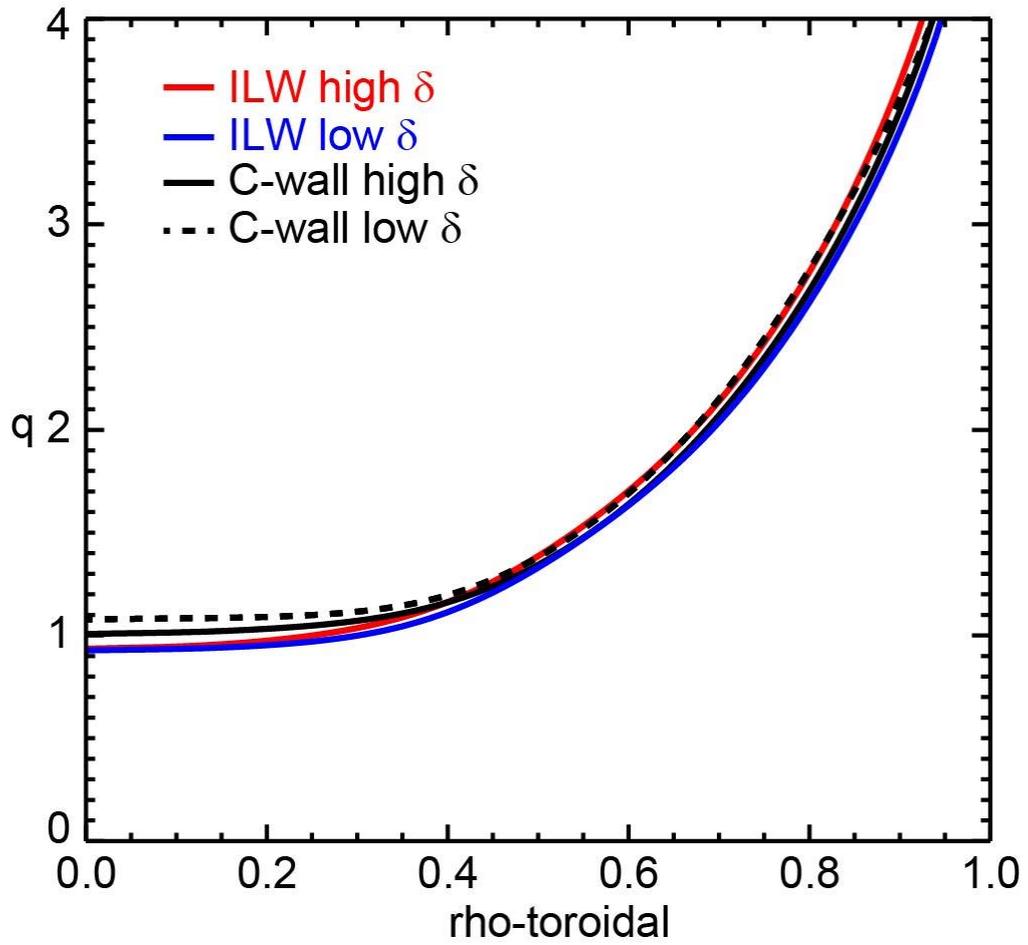

*Fig.8. q-profiles at the confinement analysis time averaged over the plasmas in each of the four power scans where MSE constrained EFIT equilibrium reconstructions are available.*

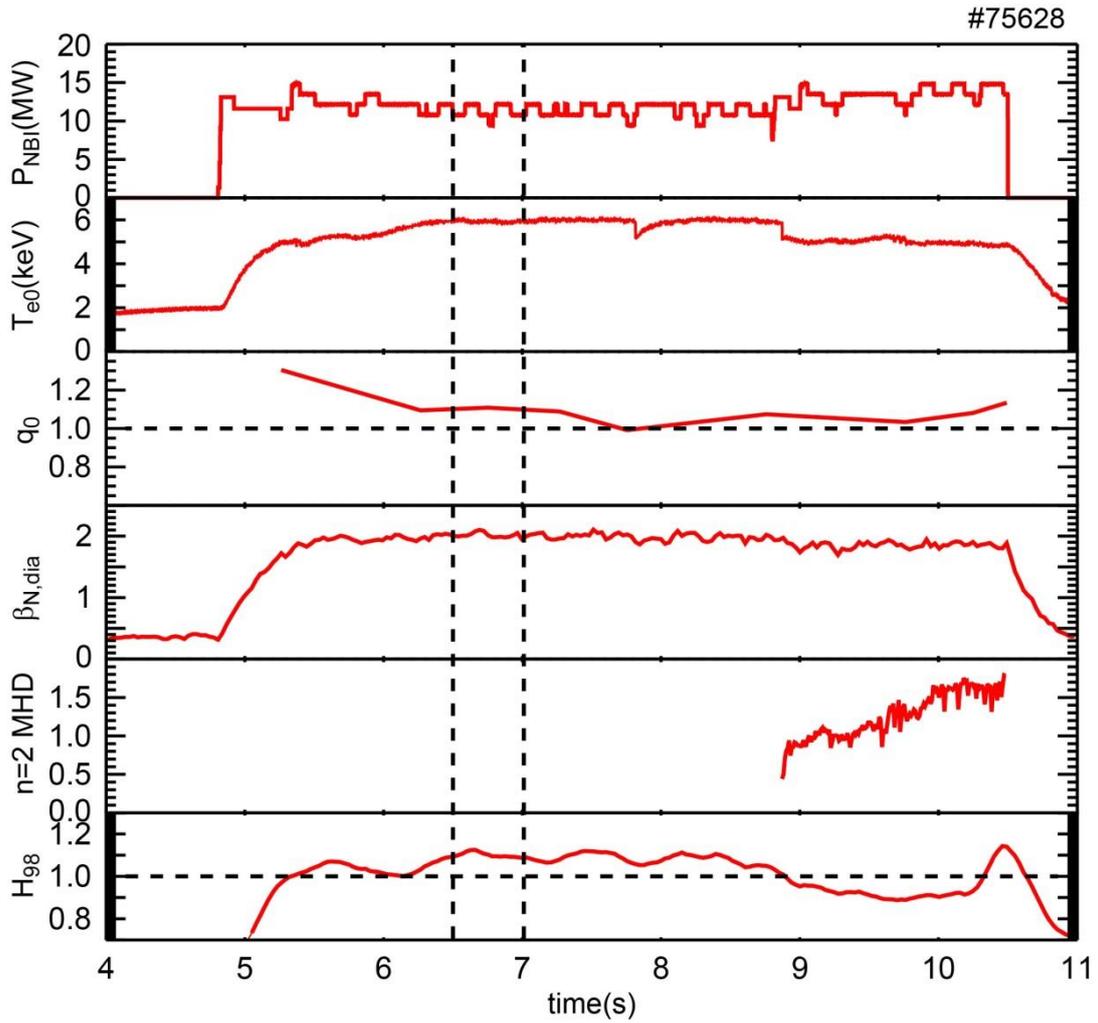

*Fig.9.Time evolution of a plasma from the C-wall low δ power scan showing the NBI power, central electron temperature from heterodyne radiometer measurements, $q_0$ from MSE constrained equilibrium reconstruction, $\beta_N$ from the diamagnetic loop measurement, the amplitude of n=2 MHD activity and $H_{98}$. The vertical lines indicate the time window used for the confinement analysis. Sawteeth are seen as abrupt drops in the central electron temperature at t=7.82s and t=8.87s.*

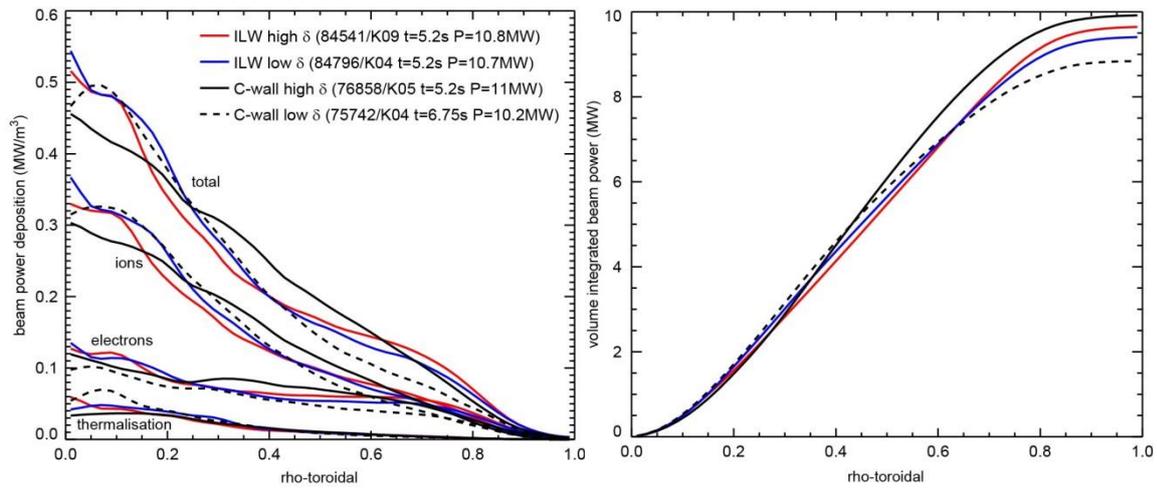

*Fig.10. Neutral beam power deposition profiles calculated using TRANSP for four plasmas with NBI heating in the range 10-11MW, one from each power scan. Profiles are shown for direct heating to the ions and electrons, the thermalisation power and the total NBI heating (left). The volume integrals of the total NBI heating power profiles are also shown (right).*

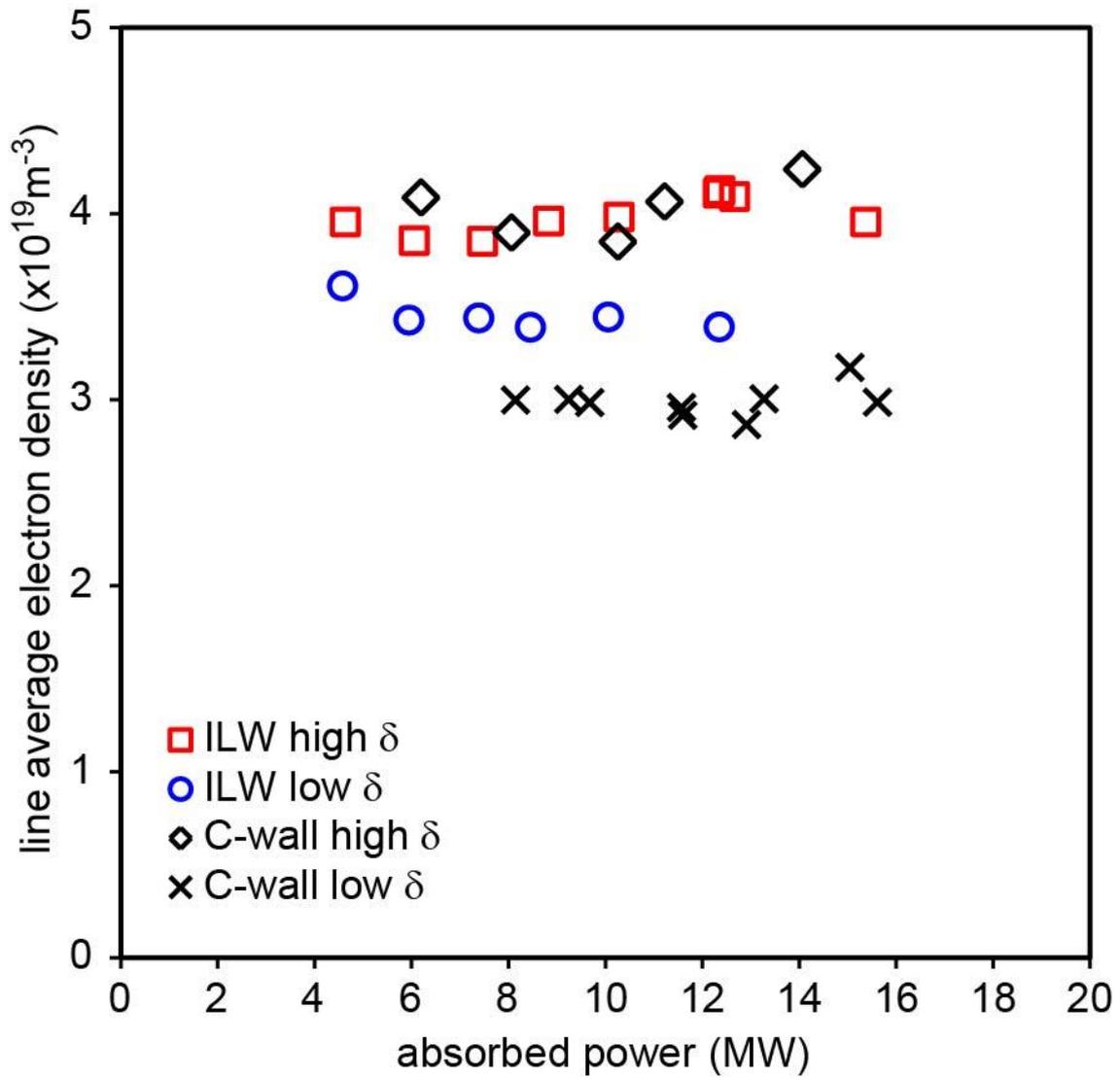

*Fig.11. Central line average electron density from high resolution Thomson scattering normalised to interferometer measurements as function of absorbed power for the four power scan experiments.*

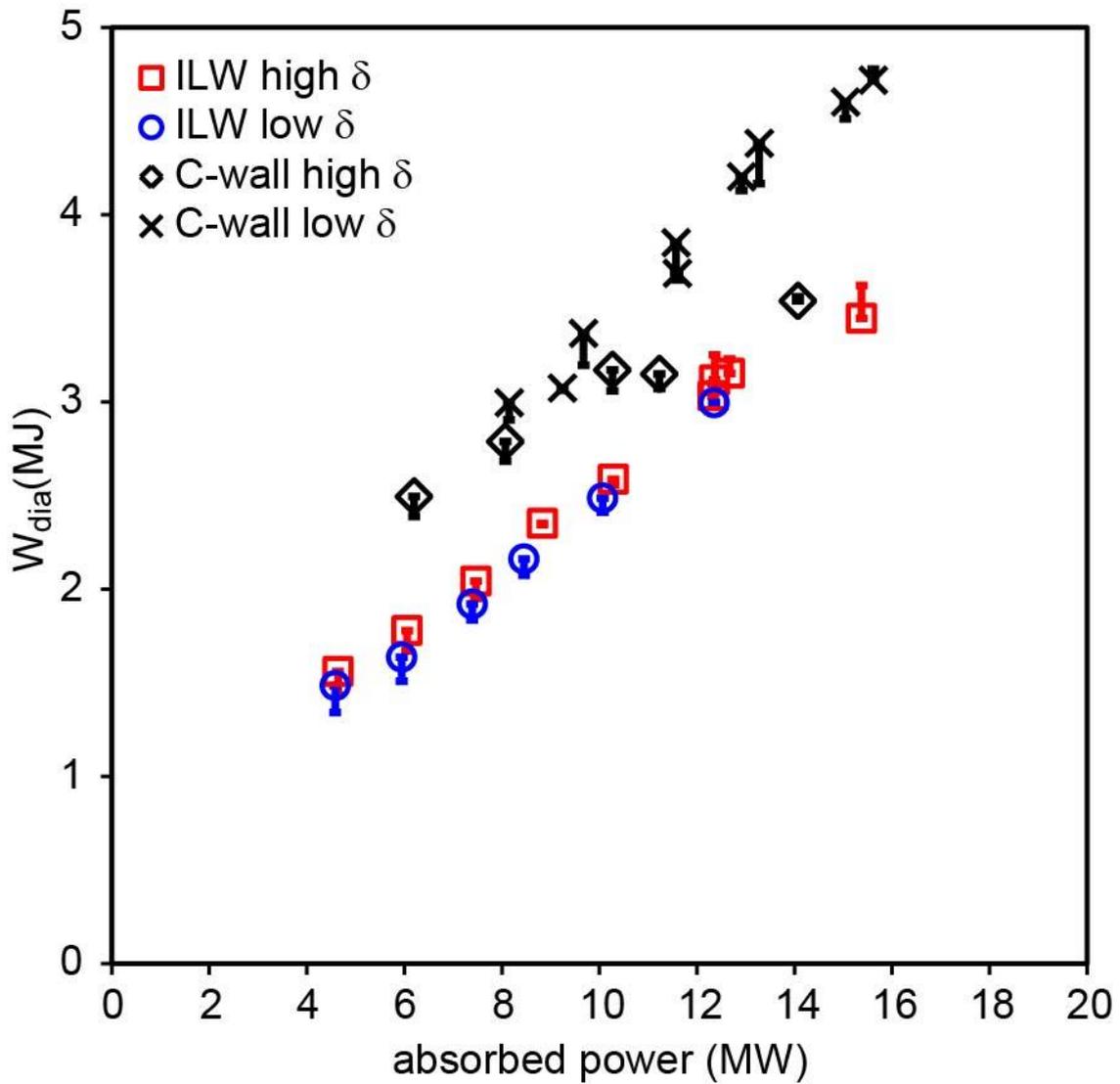

*Fig.12. Diamagnetic plasma stored energy as function of absorbed power for the four power scan experiments. The error bars indicate the discrepancy with respect to $W_{thermal}+1.5 \times W_{fast\text{-}perpendicular}$ evaluated from the measured plasma profiles using TRANSP.*

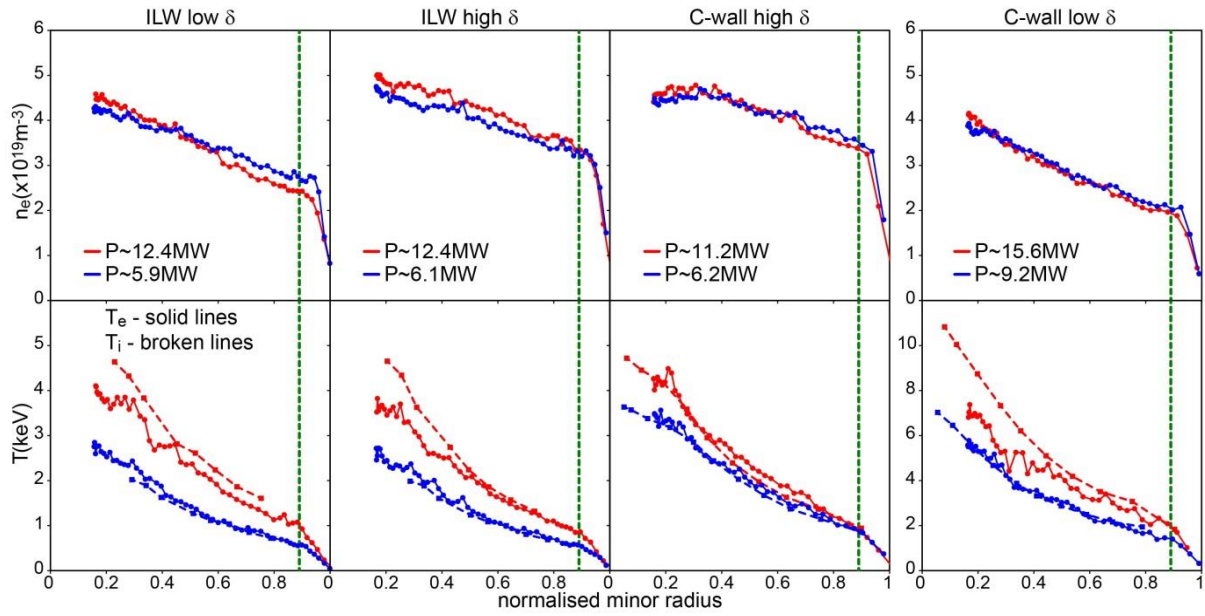

*Fig.13 Electron and ion temperature and electron density profiles at two power levels for each of the four power scans. The vertical lines indicate the boundary used for the estimation of the core and pedestal energy.*

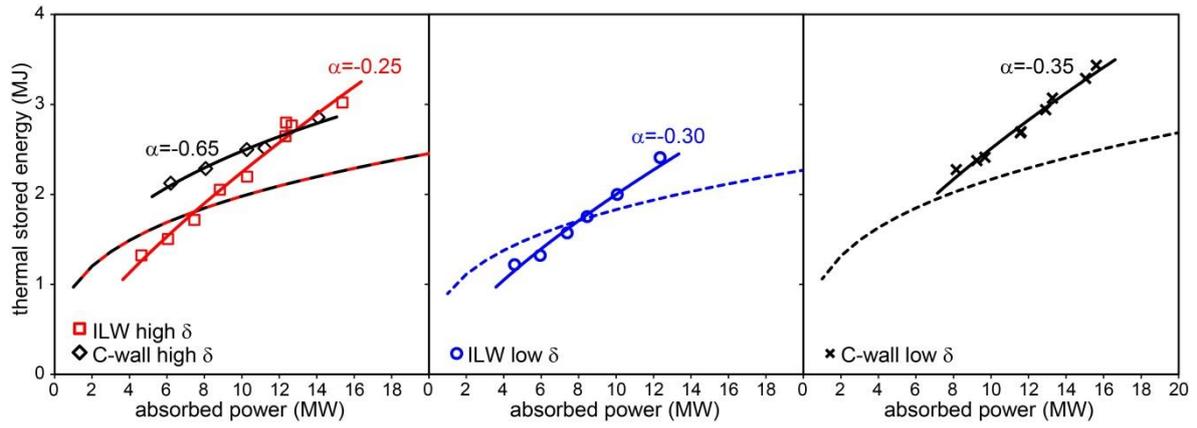

*Fig.14. Plasma thermal stored energy as a function of absorbed heating power for the four power scans. The solid lines are fits to the data assuming a scaling of the form $W_{th} \sim P^{\alpha+1}$, where α is the exponent for the scaling for energy confinement time with power. The dashed lines represent the dependence using the IPB98(y,2) scaling (i.e. α=-0.69).*

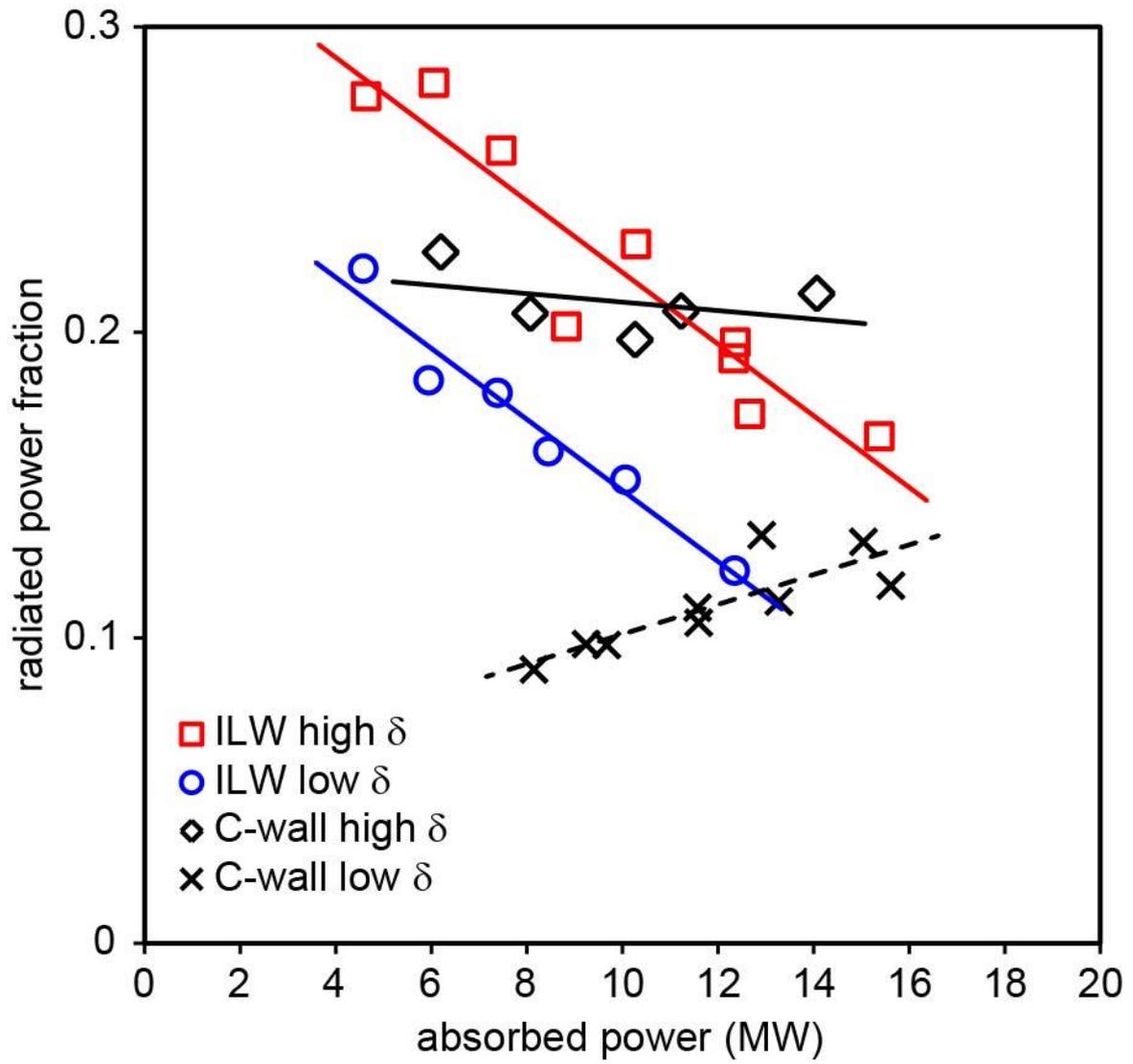

*Fig.15. Fraction of absorbed power radiated from within the last closed flux surface from bolometer measurements as a function of absorbed heating power for all four power scans. The solid and broken lines are fits to the data assuming a linear scaling.*

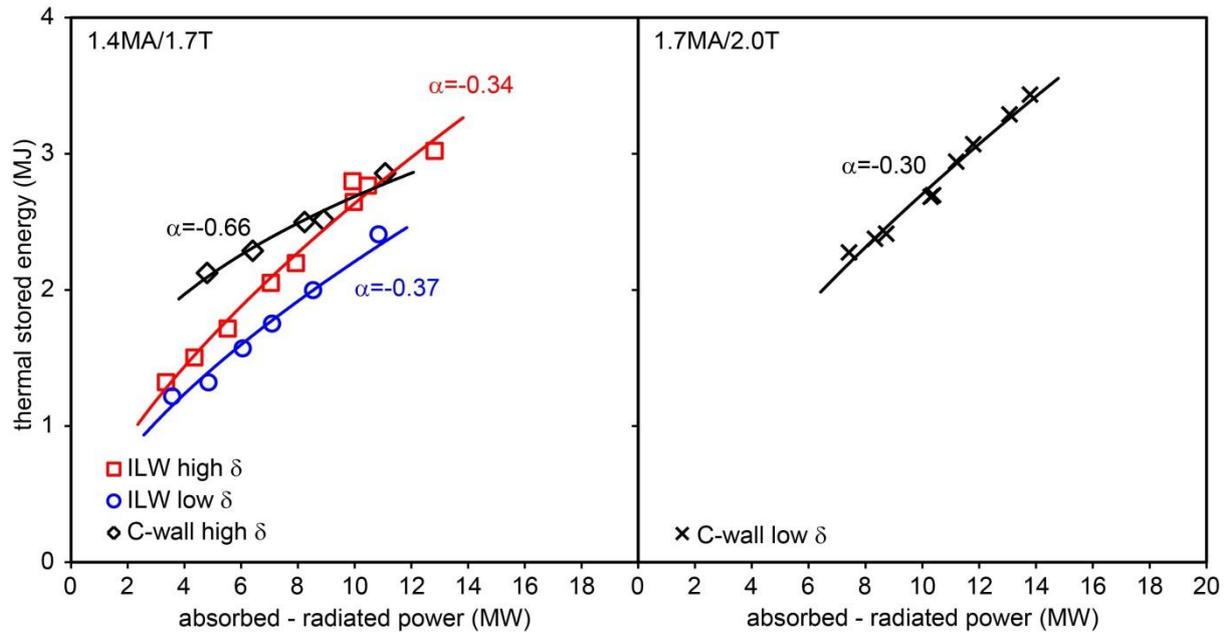

*Fig.16. Plasma thermal stored energy as a function of absorbed heating power minus power radiated from within the last closed flux surface for the four power scans. The solid lines are fits to the data assuming a scaling of the form $W_{th} \sim P^{\alpha+1}$.*

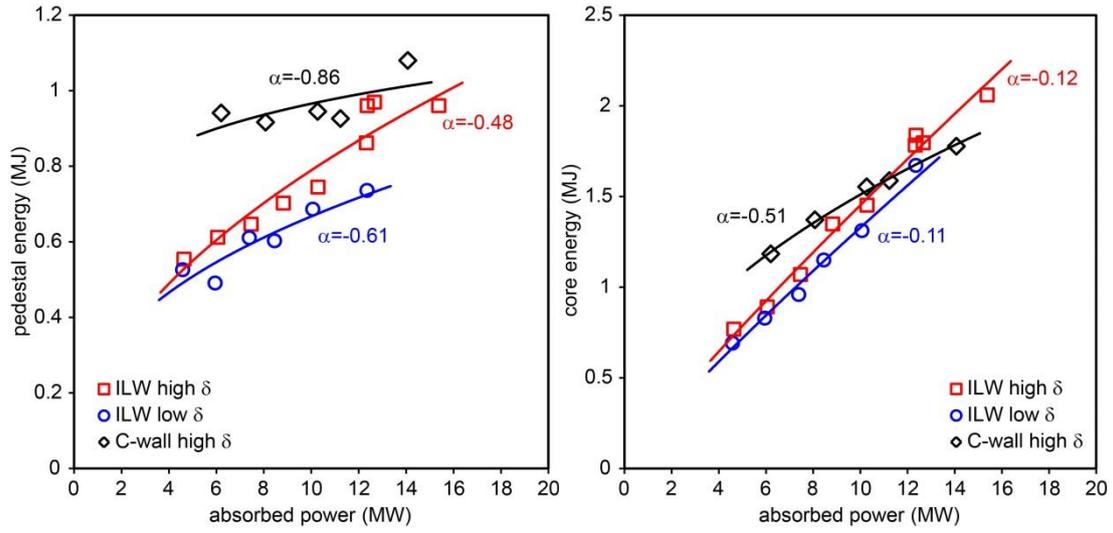

*Fig.17. Pedestal energy, $W_{ped}$, (left) and core energy, $W_{core}$, (right) as a function of absorbed heating power for three of the four power scans. The solid lines are fits to the data assuming a scaling of the form $W_{ped} \sim P^{\alpha+1}$ (left) and $W_{core} \sim P^{\alpha+1}$ (right).*

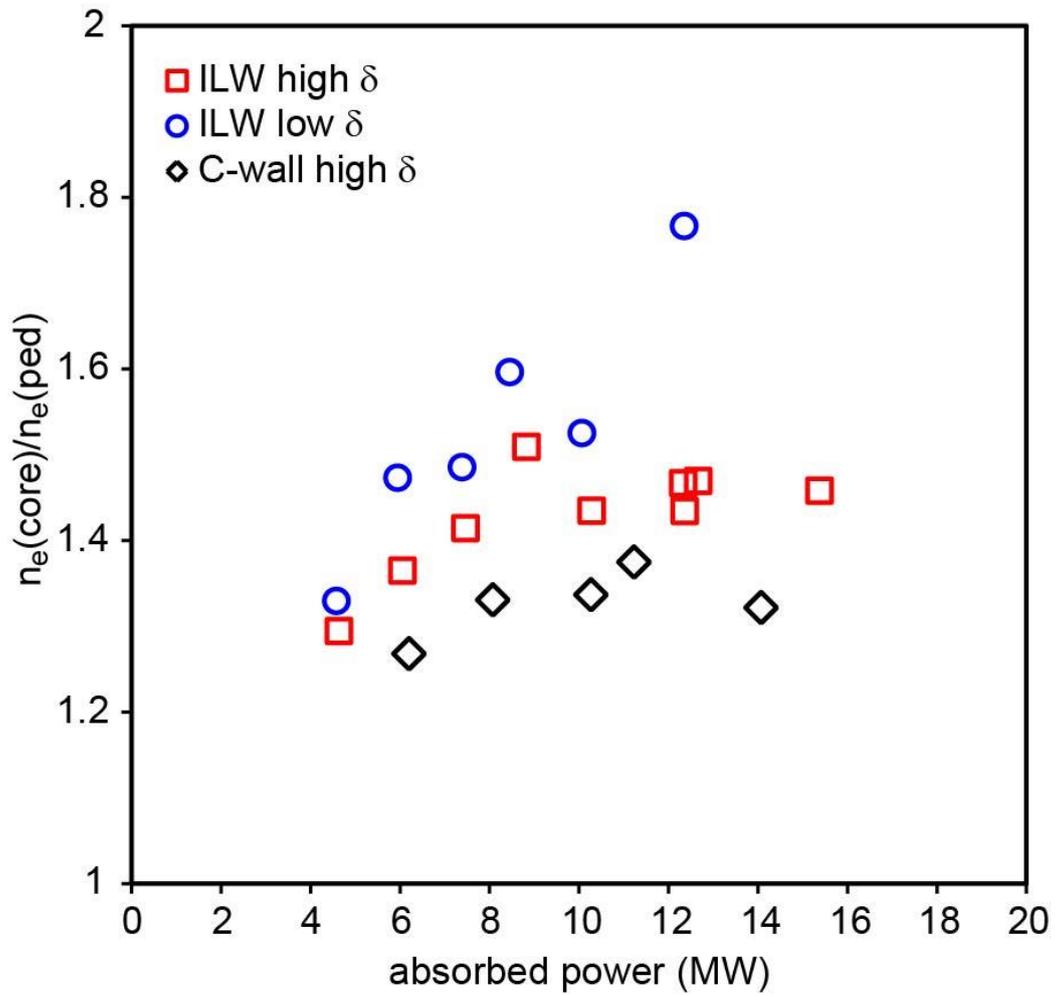

*Fig.18. Electron density peaking defined as $n_e(\rho_{tor}=0.31)/n_e(\rho_{tor}=0.89)$ as a function of absorbed heating power for three of the four power scans.*

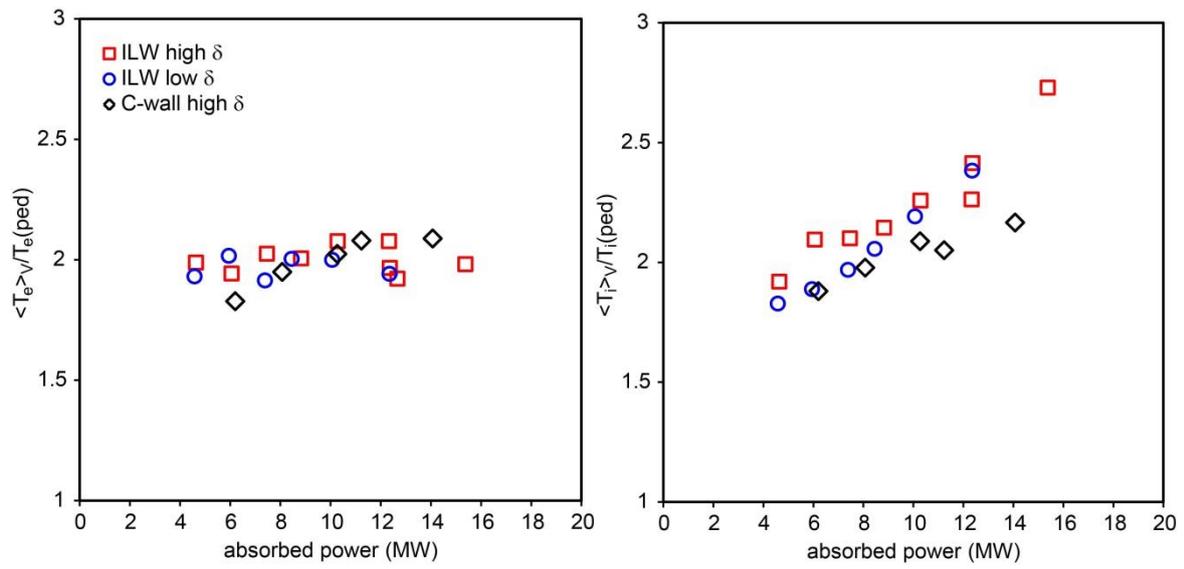

*Fig.19. Electron (left) and ion (right) temperature peaking, defined as $<T>_V/T(\rho_{tor}=0.89)$ as a function of absorbed heating power for three of the four power scans.*

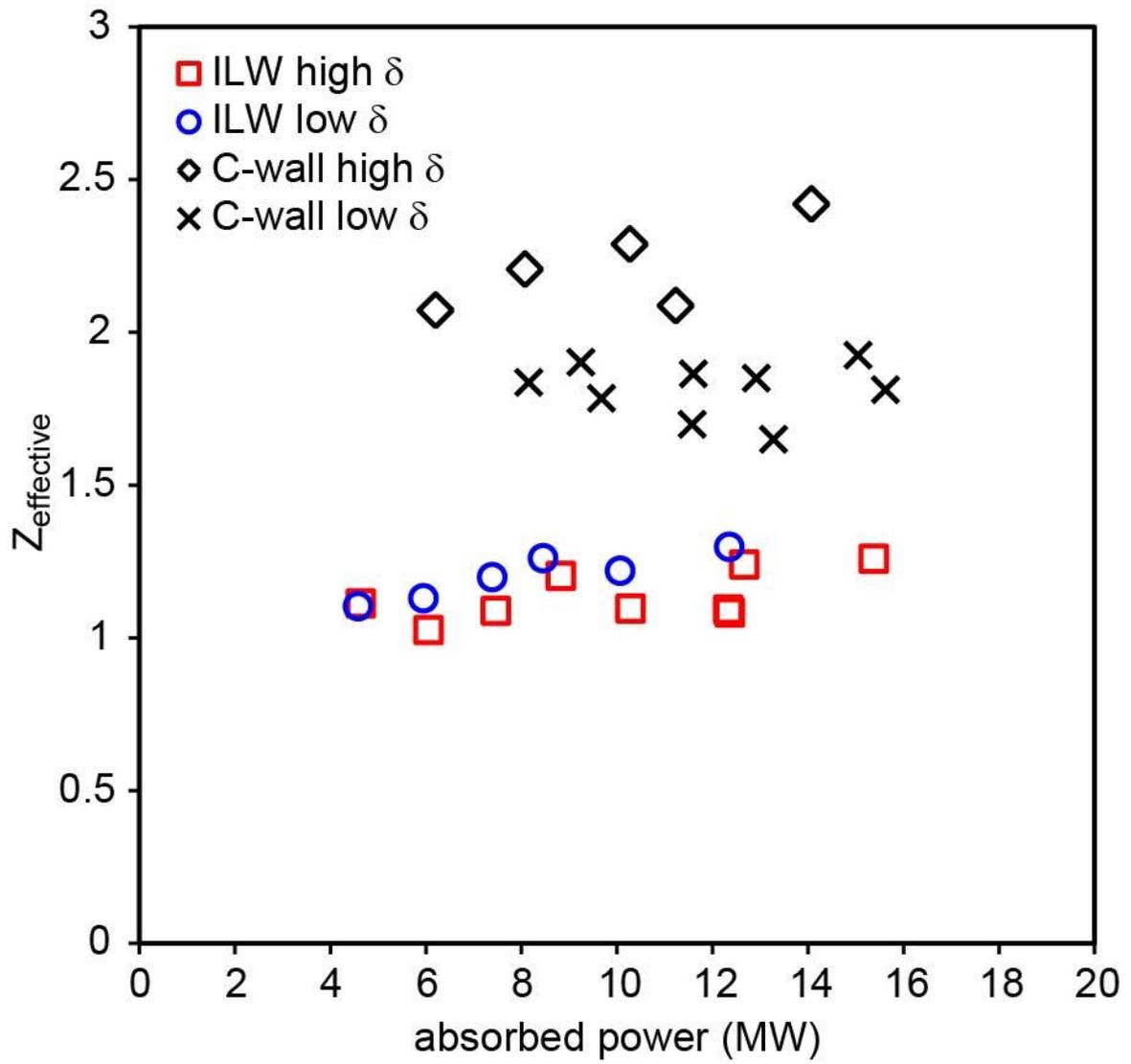

*Fig.20. $Z_{effective}$ as a function of as a function of absorbed heating power the four power scans.*

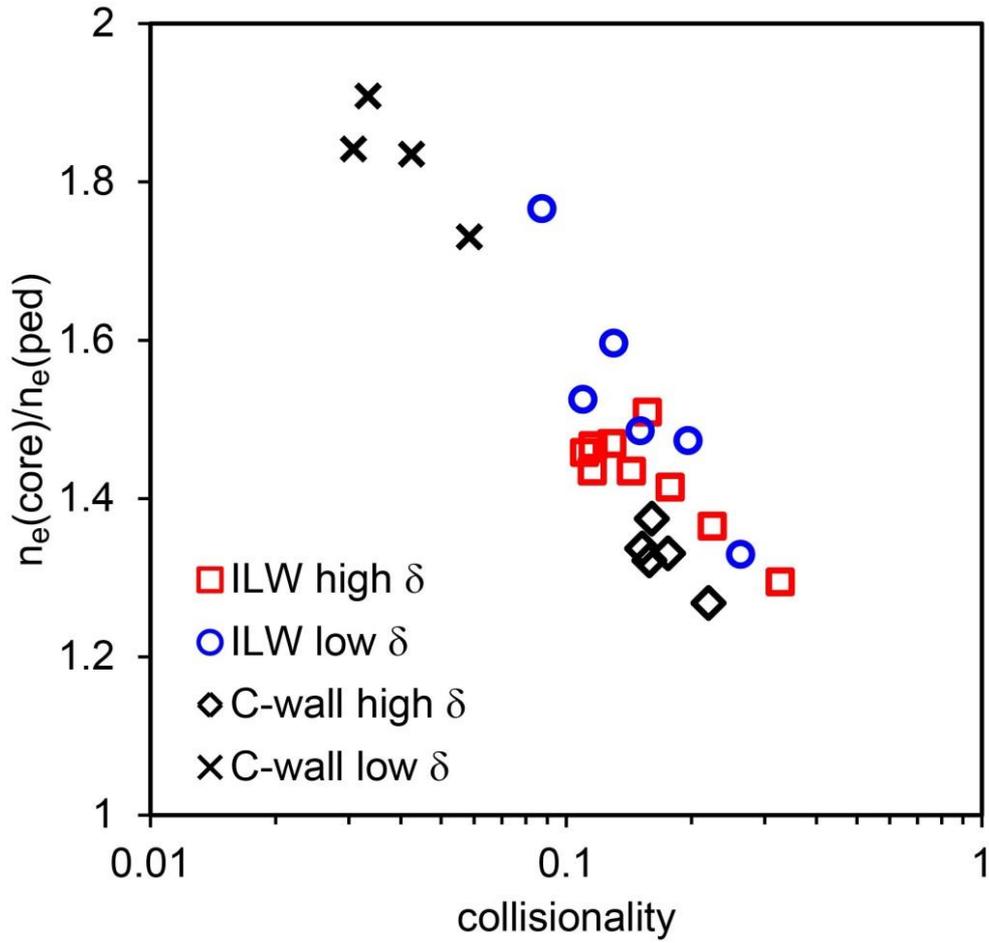

*Fig.21. Electron density peaking as a function of plasma collisionality for the four power scans. The density peaking is defined as $n_e(\rho_{tor}=0.31)/n_e(\rho_{tor}=0.89)$ and the volume averaged collisionality is defined here as $<v_{e*}>=0.012<n_e>Z_{eff}q_{95}R_0/\varepsilon^{1.5}<T_e>^2$, where density is in $10^{20}m^{-3}$, temperature is in keV and $R_0$ is the plasma geometric centre in m. Pulses without high resolution Thomson scattering data are excluded for consistency.*

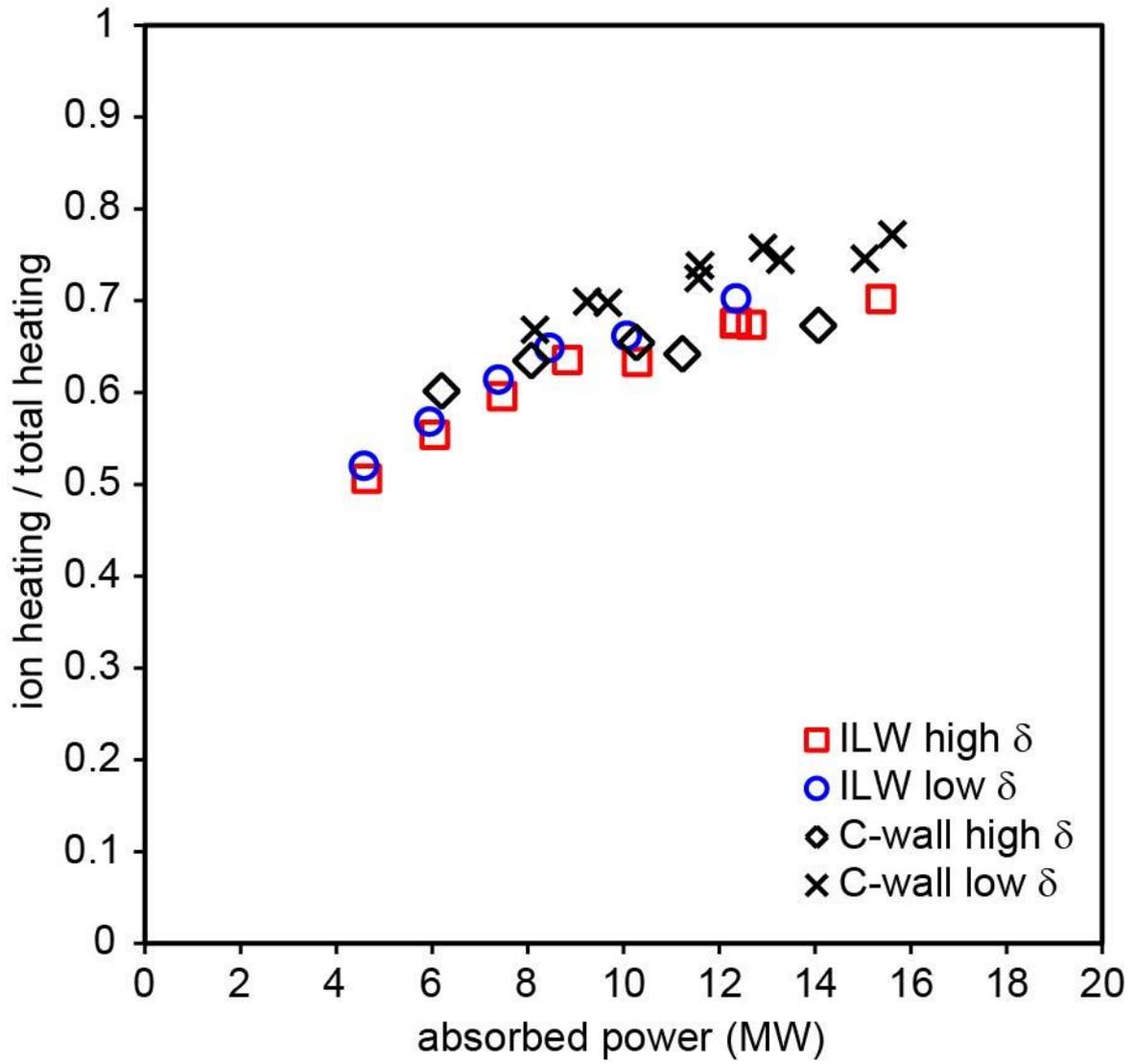

*Fig.22. Ratio of NBI ion heating to total NBI heating from TRANSP simulations as a function of absorbed heating power the four power scans.*

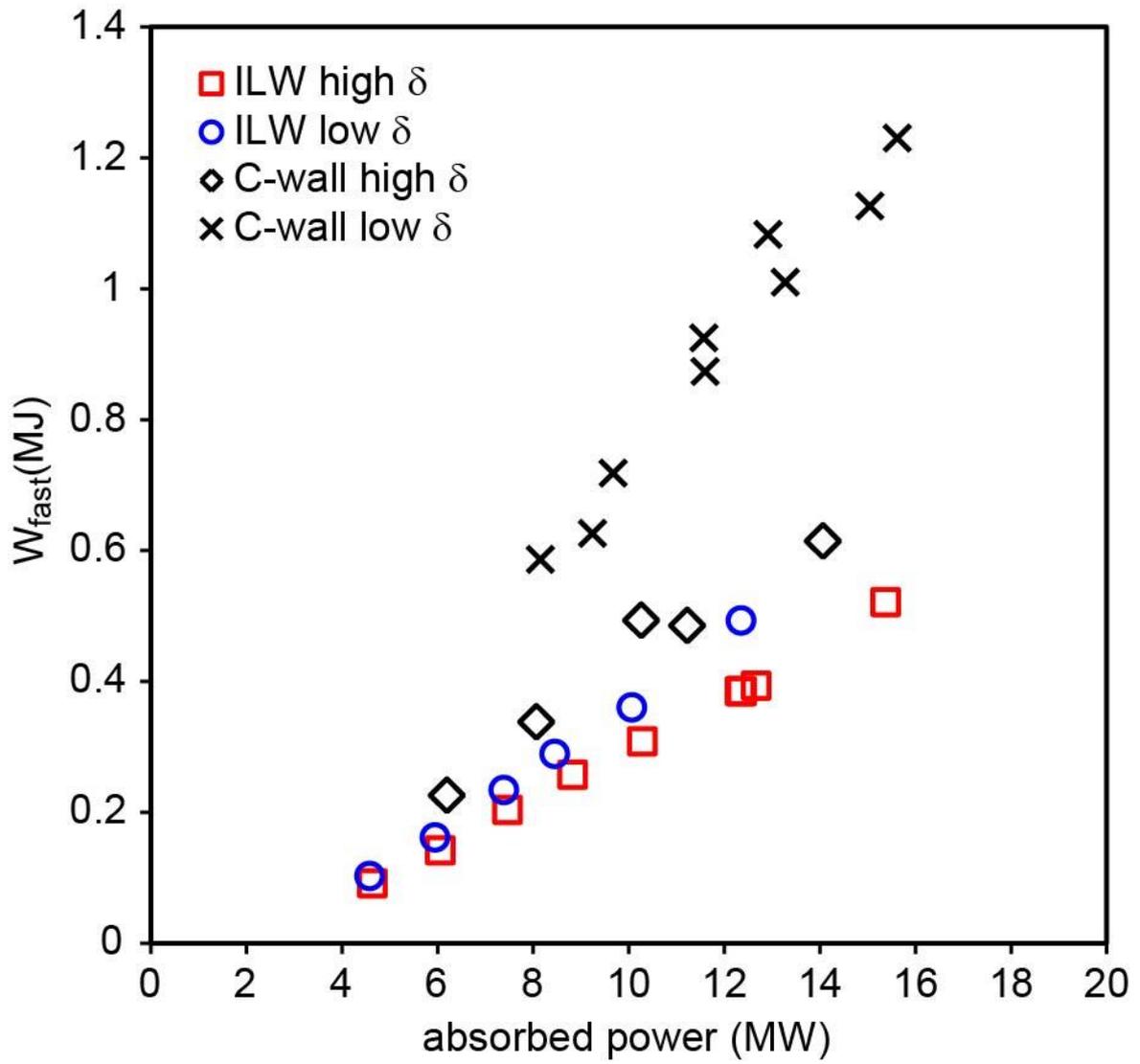

*Fig.23. Suprathermal stored energy from TRANSP NBI simulations as a function of absorbed heating power the four power scans.*

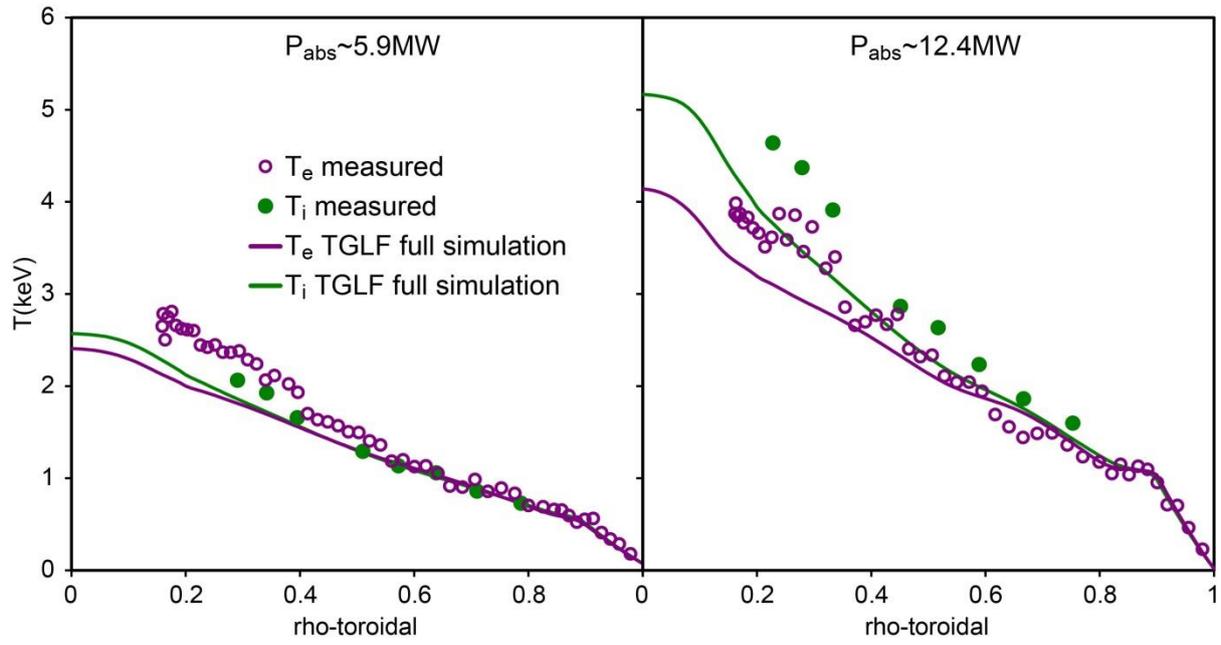

*Fig.24. Measured and modelled ion and electron temperature profiles at two power levels for the low δ ILW power scan.*

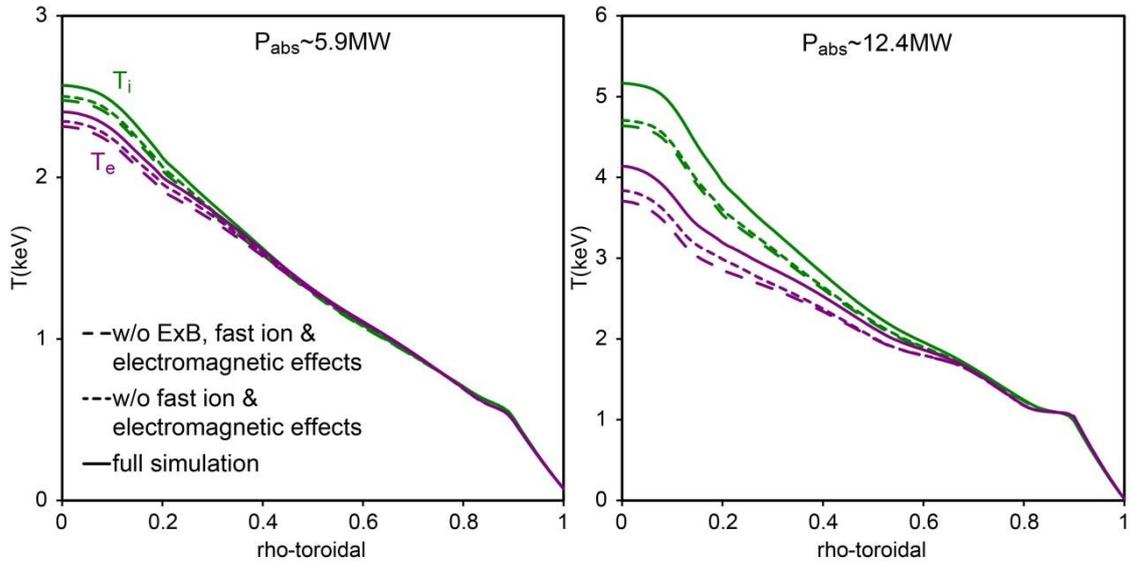

*Fig.25. Modelled ion and electron temperature profiles using TGLF at two power levels for the low δ ILW power scan. Solid lines show the full simulation, short dashes show simulations without fast ion and electromagnetic effects, and long dashes show the additional removal of E×B effects.*

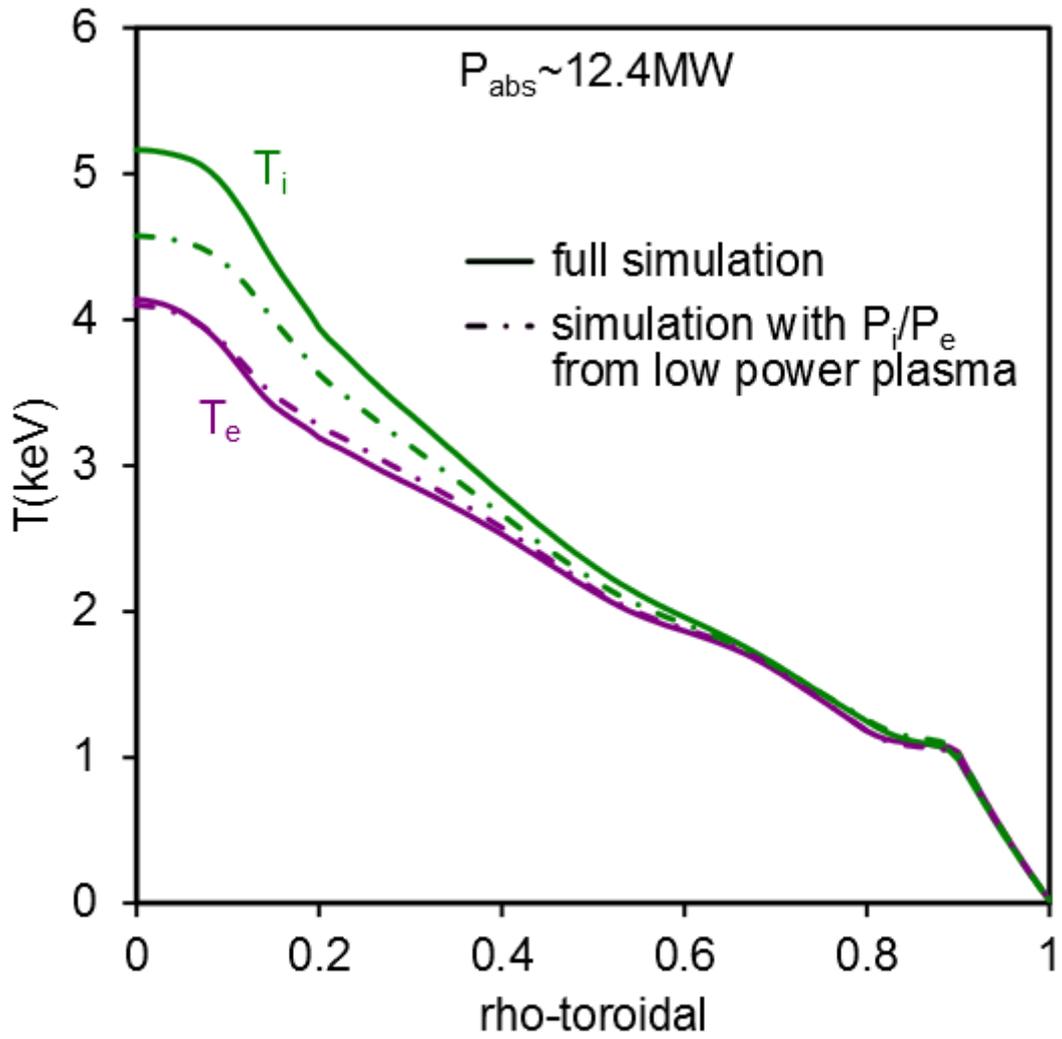

Fig.26. Modelled ion and electron temperature profiles using TGLF for a high power plasma from the low δ ILW power scan. Solid lines show the full simulation, dot-dashed lines show a simulation with the ratio of the power to the ions and electrons reduced to match that of the low power plasma in Fig.25.

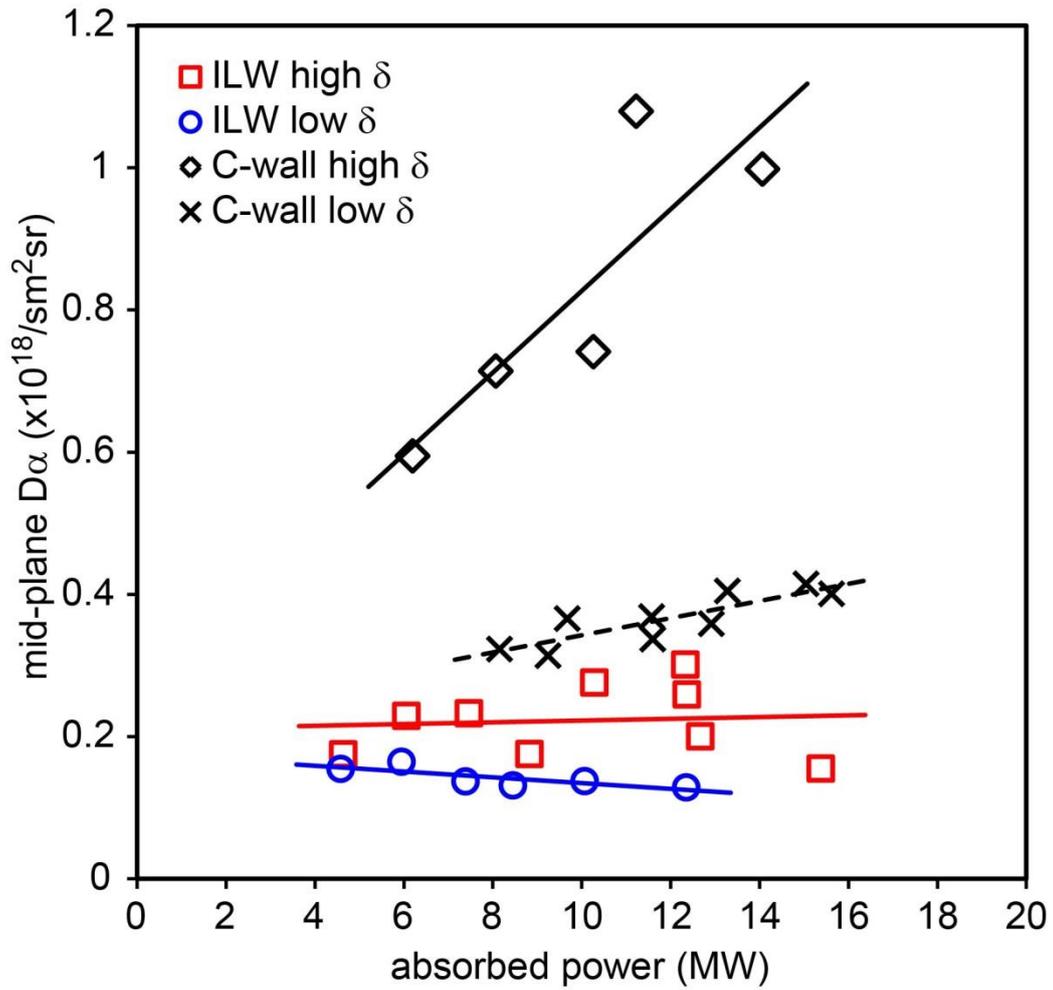

*Fig.27. Dα emission from the tokamak mid-plane line-of-sight as a function of absorbed heating power the four power scans. The level of the Dα emission is probably over-estimated for the ILW experiments due to reflections within the tokamak vessel. The solid lines are linear fits to the data.*

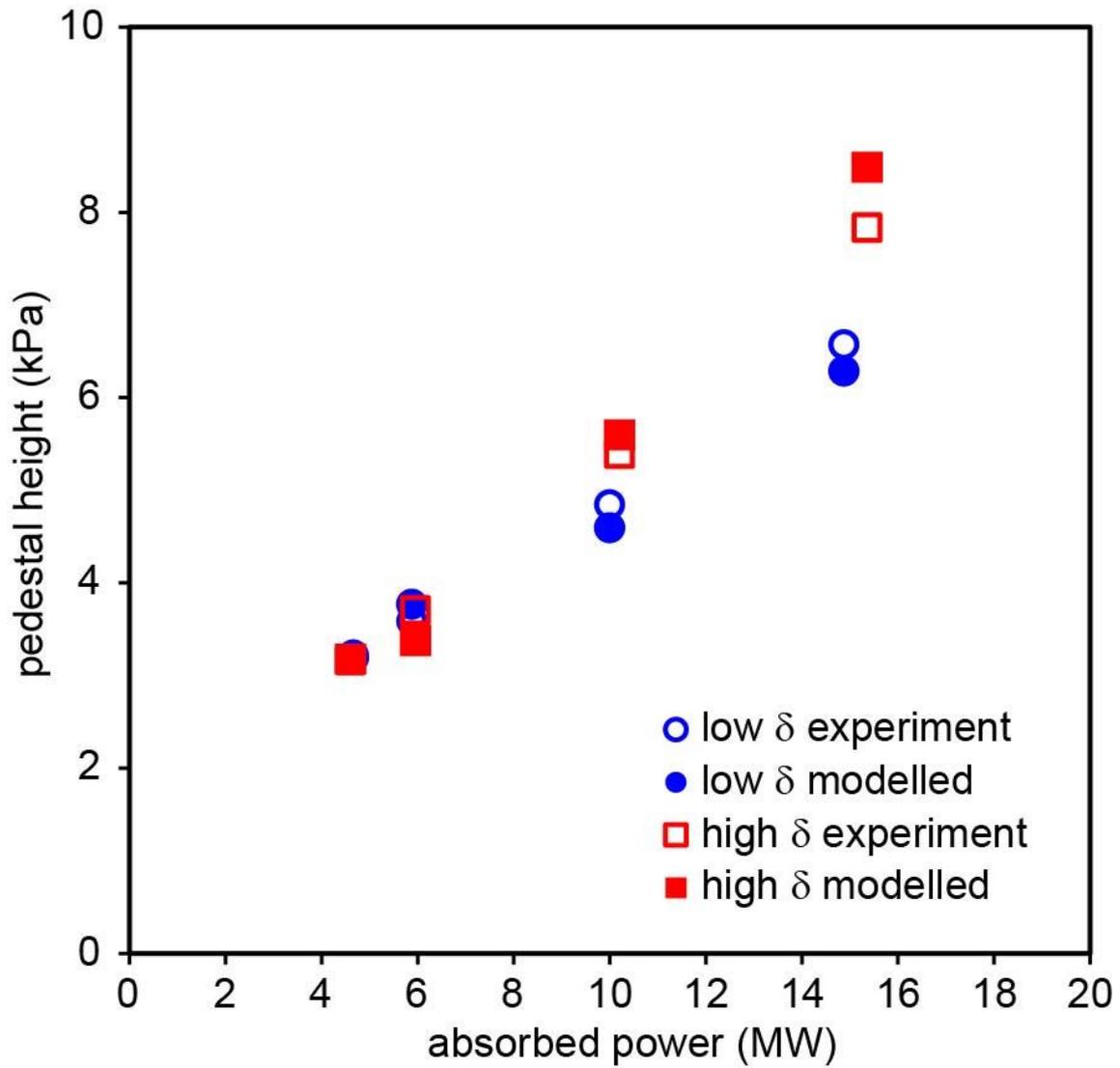

*Fig.28. Pedestal pressure as a function of heating power showing the experimental values (open symbols) and point of marginal stability from peeling-ballooning mode analysis using the ELITE code (solid symbols). The ILW high $\delta$ (triangles) and low $\delta$ (circles) are shown.*

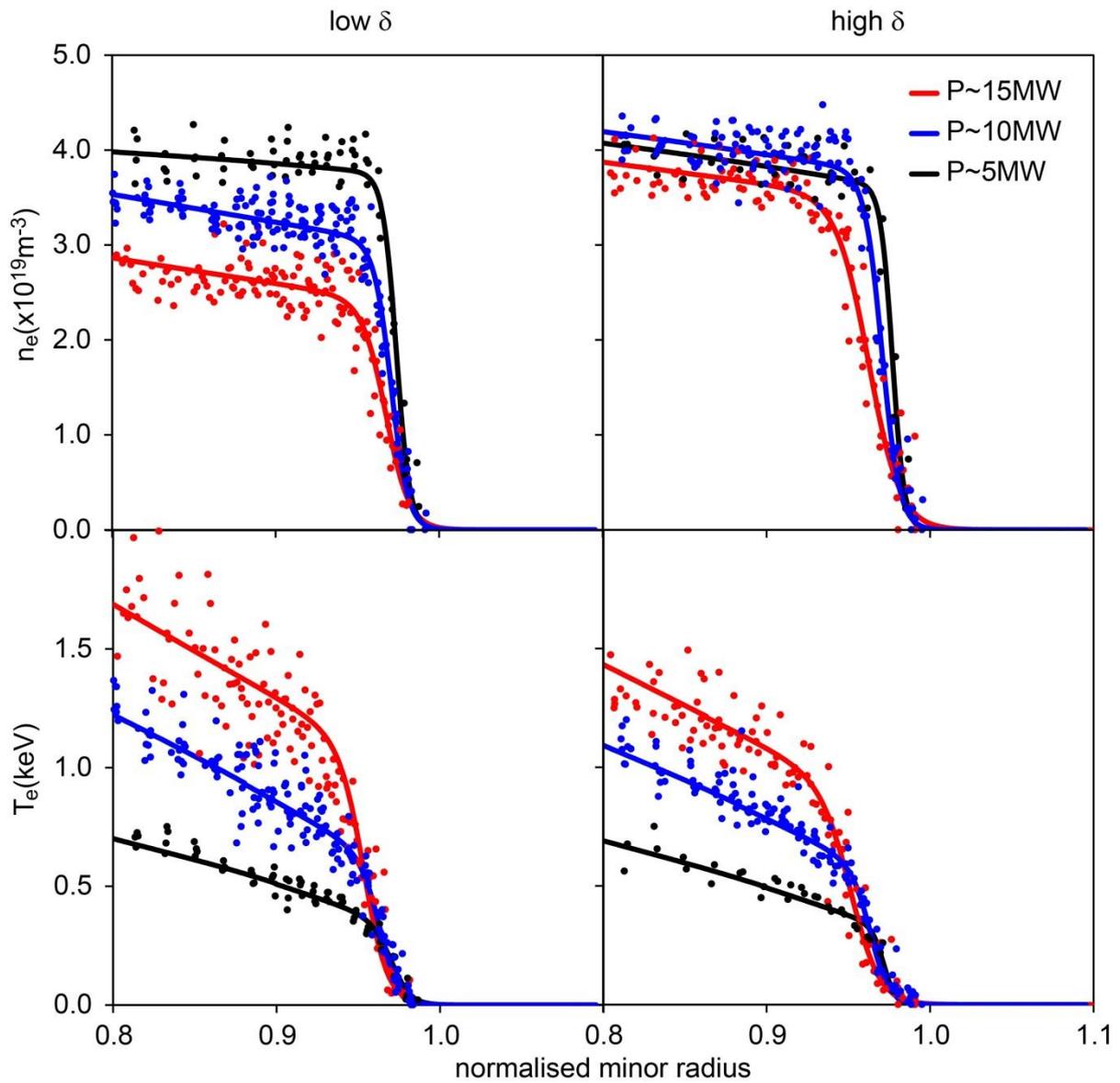

*Fig.29. Pedestal profiles for electron density and temperature for the ILW high and low δ power scans at three different power levels. The normalised minor radius has been plotted as the square-root of normalised poloidal flux. The fitted curves were used for the pedestal stability analysis.*

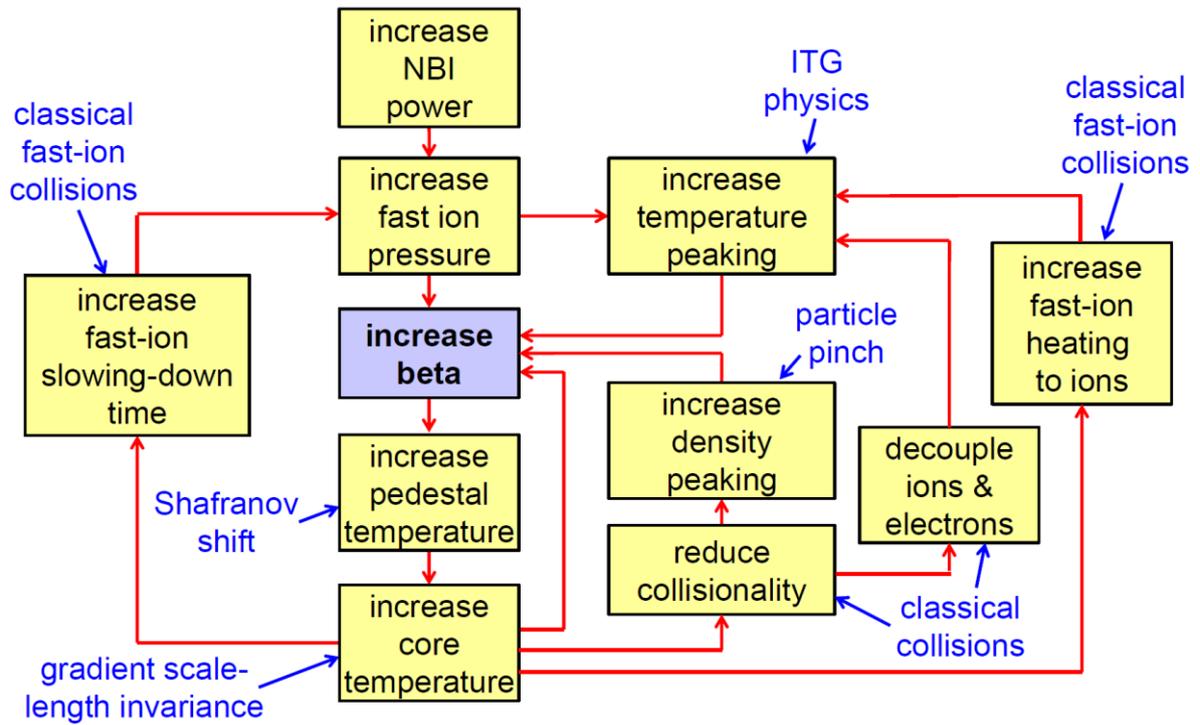

*Fig.30. Flow diagram showing some of the interdependencies suggested by the observations and modelling for the ILW power scan experiments.*

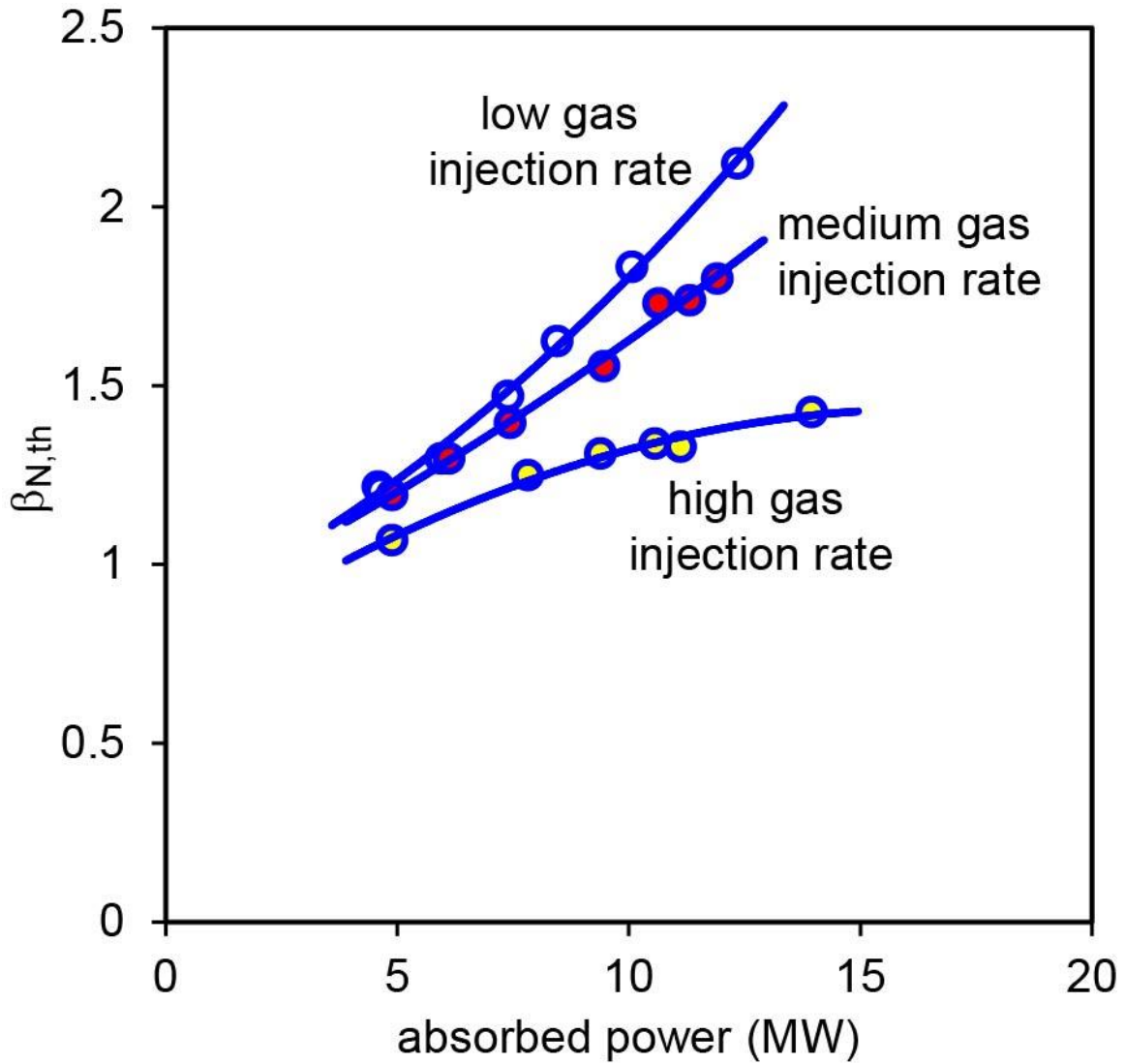

*Fig.31. Thermal component of $\beta_N$ as a function of absorbed heating power for three ILW low $\delta$ power scans: open symbols are from the power scan described elsewhere in this paper with a gas flow rate of ~$0.3\times10^{22}$ electrons/s ('low gas'); red filled symbols are with a gas flow rate of ~$0.9\times10^{22}$ electrons/s ('medium gas') and a divertor configuration optimised for pumping; and yellow filled symbols are with a gas flow rate of ~$1.8\times10^{22}$ electrons/s ('high gas') and the same divertor configuration as for 'medium gas'. The curves are polynomial fits to the data.*

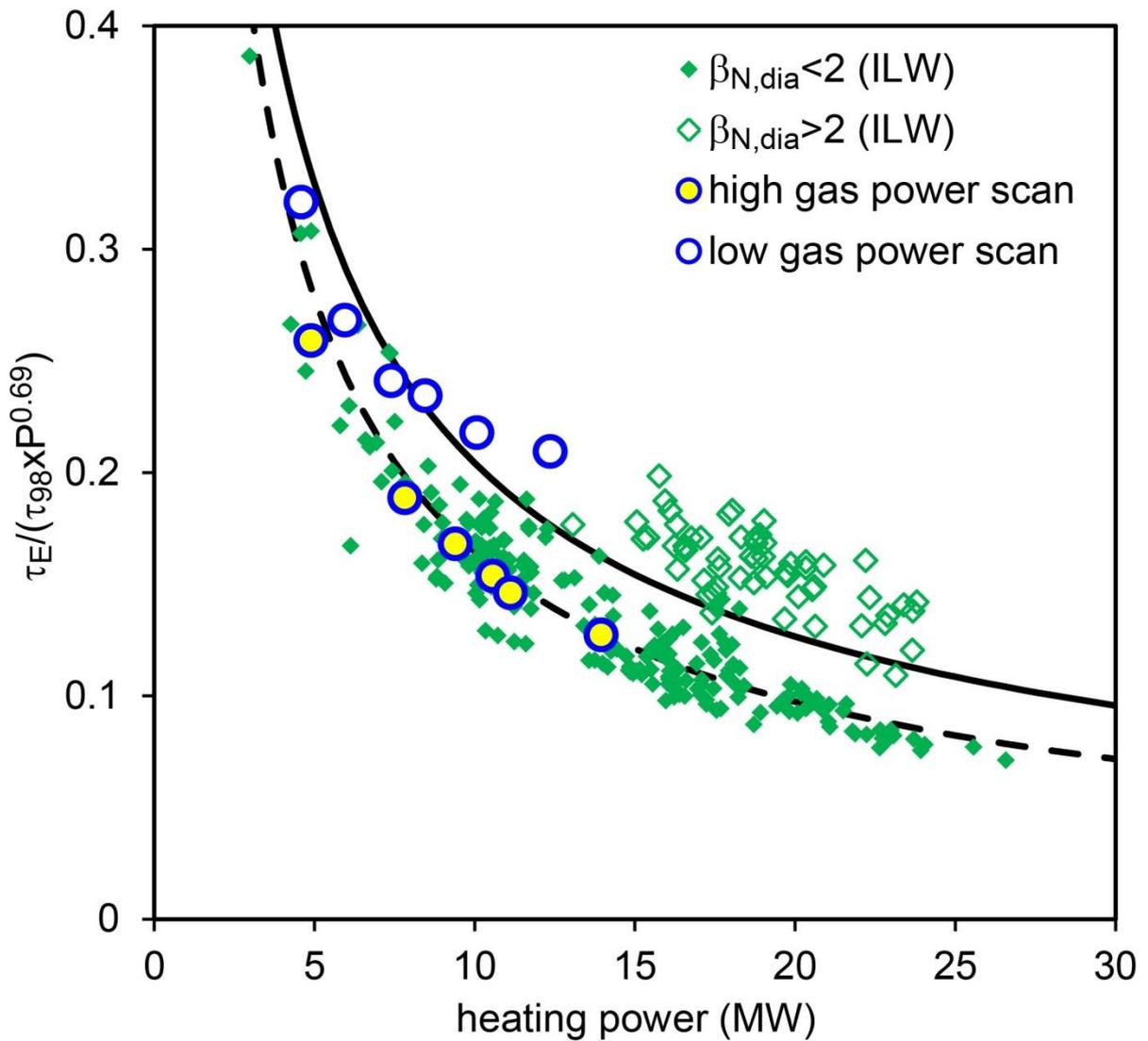

*Fig.32. Thermal energy confinement time normalised to $\tau_{98} \times P0.69$ as a function of heating power, where $\tau_{98}$ is the confinement time given by the IPB98(y,2) scaling. The JET ILW large dataset has been divided into two groups: plasmas with $\beta_{N,dia}<2$ (solid green; typical of 'baseline' experiments with high gas flow rates); and $\beta_{N,dia}>2$ (open green; typical of 'hybrid' experiments with low gas flow rates). The data from two ILW low $\delta$ power scans shown in Fig.31 are overlaid: 'low gas' (open circles); 'high gas' (yellow filled circles). The solid line represents $H_{98}=1$ and the dashed line is a fit to the $\beta_{N,dia}<2$ data.*